\documentclass[a4paper,fleqn]{cas-dc}

\usepackage[authoryear,longnamesfirst]{natbib}
\usepackage{amsmath}
\usepackage{booktabs}
\usepackage{xcolor}
\graphicspath{{figures/}}
\usepackage{pdfpages}

\ExplSyntaxOn
\cs_set:Npn \__first_footerline:
  {
    \group_begin:
    \small
    \sffamily
    \__short_authors: :~
    { \rmfamily \itshape Preprint }
    \group_end:
  }
\ExplSyntaxOff

\ExplSyntaxOn
\RenewDocumentCommand \printorcid { }
  {
    \seq_if_empty:NF \g_stm_orcid_seq
      {
        \group_begin:
          \tex_let:D \thefootnote \relax \footnotetext
            { \raggedright \textsc{orcid}(s):\c_space_token \seq_use:Nn \g_stm_orcid_seq { ;~ } }
        \group_end:
      }
  }
\ExplSyntaxOff

\newcommand{\pset}{p_{\mathrm{set}}}
\newcommand{\ptank}{p_{\mathrm{tank}}}
\newcommand{\pend}{p_{\mathrm{end}}}
\newcommand{\pneed}{p_{\mathrm{need}}}
\newcommand{\pref}{p_{\mathrm{ref}}}
\newcommand{\pdiff}{p_{\mathrm{diff}}}
\newcommand{\Qcomp}{Q_{\mathrm{comp}}}
\newcommand{\Qtank}{Q_{\mathrm{tank}}}
\newcommand{\Qspec}{Q_{\mathrm{spec}}}
\newcommand{\Vtank}{V_{\mathrm{tank}}}

\begin{document}
\let\WriteBookmarks\relax
\def\floatpagepagefraction{1}
\def\textpagefraction{.001}

\shorttitle{Retrofitting centralized air compressor stations with end-use-aware predictive set-pressure control}
\shortauthors{T. Watanabe et~al.}

\title[mode = title]{Retrofitting centralized air compressor stations with end-use-aware predictive set-pressure control using a physics-constrained pressure-drop model}

\author[1]{Taito Watanabe}
\credit{Conceptualization, Methodology, Software, Validation, Investigation, Data curation, Writing -- original draft}

\author[1]{Yohei Kono}
\cormark[1]
\ead{yohei.kono.un@hitachi.com}
\credit{Conceptualization, Methodology, Writing -- review \& editing, Supervision}

\author[2]{Akira Iyozumi}
\credit{Resources, Investigation, Writing -- review \& editing}

\author[2]{Takashi Nakajima}
\credit{Resources, Investigation, Writing -- review \& editing}

\affiliation[1]{organization={Research \& Development Group, Hitachi, Ltd.},
            addressline={292 Yoshida-cho, Totsuka-ku},
            city={Yokohama},
            postcode={244-0817},
            state={Kanagawa},
            country={Japan}}

\affiliation[2]{organization={Global Air Power Group, Hitachi Industrial Equipment Systems Co., Ltd.},
            addressline={390 Muramatsu, Shimizu-ku},
            city={Shizuoka},
            postcode={424-0926},
            state={Shizuoka},
            country={Japan}}

\cortext[1]{Corresponding author}

\begin{abstract}
Compressed air systems usually run at a constant set pressure well above the end-use requirement; the excess margin is wasted energy. A lightweight hierarchical nonlinear model predictive control (NMPC) scheme, retrofitted to the sequencing controller of a centralized compressor station typical of small and medium-sized plants, lowers the set pressure as far as the end-use requirement allows. It uses only two pressure gauges (receiver and one critical end-use point) and compressor data, without flow meters or piping surveys. The end-use pressure is predicted by a low-order pressure-drop model identified by least squares under stability, damping and steady-state equilibrium constraints, which serve use rather than accuracy: they let the measured pressure drop be held over the horizon with unit gain. Validated in three stages (plant data, a pipe-network simulator checked against measurements, and the actual controller, at most 1.41~s per update), the NMPC meets the required pressure under real demand and reduces the estimated annual energy consumption by 2.8\,\% relative to the current constant operation (9.3\,\% relative to 0.70~MPa). A PID cascade with the same gauge saves about as much only when tuned by a standard rule at its recommended setting from a constant-load step test, difficult in an operating plant; the NMPC needs no such test and keeps its performance with its initial model and default weights after plant expansion and network split. The margin becomes a design parameter instead of a fixed excess, and the controller's signals show where the supply system lacks headroom.
\end{abstract}


\maketitle
\hypersetup{pdfauthor={Taito Watanabe, Yohei Kono, Akira Iyozumi, Takashi Nakajima},
  pdfsubject={Preprint},
  pdfkeywords={Compressed air systems, Model predictive control, Physics-constrained system identification, Set-pressure optimization, Industrial controller implementation, Energy efficiency}}

\section{Introduction}\label{sec:intro}

\subsection{Background and practical constraints}\label{subsec:background}
Compressed air systems account for about 10\,\% of industrial energy use in several countries \citep{Saidur2010}, and electricity is by far the largest cost of owning and operating them \citep{DOE_CA_Guide}. Energy saving has so far focused on more efficient compressors and on sequencing controllers that coordinate several units \citep{Saidur2010,DOE_CA_Guide}. Further gains require operating the whole system, from the compressors through the pipe network to the end-use equipment. Lowering the supply pressure reduces power directly: a rule of thumb gives about 1\,\% per 2~psi (0.014~MPa) \citep{DOE_CA_Guide}, and a compressor manufacturer reports 8.4\,\% for lowering the discharge pressure of a 75-kW single-stage screw compressor at full load from 0.69 to 0.59~MPa \citep{HitachiIES_energy}. Yet most plants run their compressors at a constant set pressure far above what the end-use equipment requires, as a guard against pressure drops caused by sudden air consumption. The excess margin during low-demand periods is wasted energy.

Reducing this margin, that is, lowering the set pressure only as far as the end-use requirement allows, is simple in principle, but four practical constraints apply. First, pipes are often hidden in ceilings and under floors, so surveying diameters, lengths and connections to build a physical model is costly. Second, flow meters cost much more than pressure gauges, so installing them at many end-use points is not cost-effective. Third, the control law must run on the industrial controller already installed in the sequencing controller, which has limited computing power and memory. Fourth, piping is frequently added or removed as products and equipment change, so the system dynamics drift over time. A fixed-gain controller tuned at commissioning therefore degrades, leading either to unstable pressure or to a return to conservative margins. This study addresses plants in which one centralized compressor station feeds the whole network, the default layout in industrial practice \citep[Sect.~3.5.1]{AtlasCopco_CAM9} and typical of small and medium-sized factories; sites whose consumers are spread over several buildings are instead supplied by satellite compressor rooms or decentralized compressors \citep[p.~38]{DOE_CA_Guide}, each of which forms such a system.

\subsection{Existing approaches and their limitations}\label{subsec:related}
System-level approaches fall into three groups: (a) optimization using detailed physical models or numerical simulators of the surveyed piping \citep{Krichel2014,Friedenstein2018}; (b) measures at the individual end-use points, such as local pressure regulation and monitoring of air consumption \citep{Grybos2024}; and (c) data-driven prediction and optimization. Examples of (c) include neural-network load forecasting of compressors \citep{wu_air_compressor}, fuzzy--neural adaptive control of vacuum pressure \citep{Sanjaya2025FuzzyLA}, polynomial regression with metaheuristics for set-pressure optimization \citep{mallia2024intelligent}, model-free adaptive control \citep{hou2013survey} and nonlinear model predictive control (NMPC) based on neural-network models \citep{piche2000nonlinear}. Model predictive control (MPC) has also been applied on the compressor side rather than to the pressure at a critical end-use point: to the discharge pressure of a variable-speed screw compressor, in comparison with PI control \citep{Quartarone2013}, and to the unit commitment and operating points of a compressed air network using demand forecasts \citep{Ravasio2023}.

Measured against the constraints above, these approaches fall short in three respects. \textbf{Scope}: most methods in (c) address a single compressor or a local process and ignore pipe-network pressure drops and the interaction among end-use points. \textbf{Cost}: (a) requires a piping survey and (b) requires instruments and communication at every end-use point, which violates the first two constraints; none of (a)--(c) offers a procedure for facility changes other than manual remodeling or retuning. \textbf{Physical consistency}: purely data-driven models in (c) guarantee neither mass conservation nor stability, so multi-step predictions can diverge or lose accuracy outside the training range. Grey-box identification with prior physical knowledge \citep{Tulleken1993} and identification with stability constraints \citep{Lacy2003,Miller2013} exist, but they have not been applied to end-use pressure prediction in compressed air systems or evaluated together with a controller implementation; identification for predictive control has instead used multi-step-ahead criteria \citep{Shook1991,Gopaluni2004}.
 Operational optimization of compressed air systems \citep{Zhong2023,Lassahn2026} and MPC of the pressure in industrial gas headers \citep{Wiid2021} have also been reported, and pressure controllers for water distribution networks have been designed from identified models and evaluated on detailed network simulators \citep{Galuppini2021}; these designs do not simultaneously minimize instrumentation, run on existing controllers and cope with facility changes.

\subsection{Contributions}\label{subsec:contribution}
This paper accepts the four constraints as design requirements. It proposes a hierarchical NMPC that lowers the set pressure while meeting the end-use requirement, and validates it in stages on plant data, a simulator checked against measurements and the actual sequencing controller. The contributions are as follows.
\begin{enumerate}[1.]
\item \textbf{End-use-aware control with one added pressure gauge.} The set pressure is lowered against the end-use requirement using only the receiver and end-use pressure gauges and the discharge flow converted from compressor data, so the only added instrument is one gauge at the critical end-use point and neither a piping survey nor flow meters are needed, removing the first two constraints of Section~\ref{subsec:background}. The method is compared with an end-use pressure PID cascade that needs the same added gauge, the alternative a practitioner could install at the same cost, so that the benefit of the controller is separated from that of the added measurement.
\item \textbf{Physics-constrained identification of an end-use model usable for prediction.} The end-use pressure model, a low-order autoregressive model with exogenous input (ARX) written in terms of the receiver-to-end-use pressure drop, is identified by least squares under constraints on damping, stability and a steady-state pressure-drop equilibrium; the receiver and compressor models need only the rated volume and one set-pressure change per compressor product. The constraints build the known pressure response of a pipe network into the data-driven model, which makes it physically consistent in the way the NMPC uses it: with the measured pressure drop held over the prediction horizon, the prediction is stable and the held quantity is not amplified, whereas unconstrained identification with the same regressors diverges in this use. The model is obtained from operating data by a fixed procedure, without a piping model.

\item \textbf{Control without plant-specific tuning that is robust to facility changes.} Unlike a PID cascade with the same instrumentation, whose tuning needs a constant-load step test in the running plant and a choice of the closed-loop time constant, the controller uses fixed default weights, needs no such test and is not retuned after piping or receiver changes, so the set-pressure reduction is kept instead of falling back to the conservative margins described in Section~\ref{subsec:background}. This robustness comes from the model structure rather than from tuning: the end-use model is written in terms of the pressure drop and has a constant-drop equilibrium, so a facility change that shifts the pressure-drop level is absorbed by the online estimate of the pressure drop, and only a change in the dynamics themselves calls for rerunning the same constrained identification.
\item \textbf{Lightweight NMPC on an existing controller.} With a single decision variable, analytic Jacobians and a fixed iteration limit, the NMPC runs on the same hardware as existing sequencing controllers, with a computation time bounded well below the compressor response time.
\end{enumerate}
ARX models of the pressure dynamics through a compressed air network were first identified by one of the authors to estimate air consumption and predict the end-use pressure, without control \citep{Kono2024SICE,Kono2024DFD}. This paper extends the authors' conference paper \citep{Watanabe2026rengo}, which introduced the modeling method and the NMPC structure (contributions 2 and 4 in preliminary form) and reported one-step accuracy on plant data, a comparison with constant set-pressure operation under synthetic load, and the execution time on the controller. The new material, which supports contributions 1 and 3, is the comparison with the PID cascade and its tuning sensitivity, closed-loop evaluation with real demand at the controlled end-use point, plant-expansion and network-split scenarios, the multi-step comparison with unconstrained identification, comparison of the pipe-network simulator with measurements, and analysis of the execution-time distribution. The compressor model was re-identified, the end-use prediction errors were recomputed and the simulation conditions were updated, so the numerical results differ from those in the conference paper.

\subsection{Staged validation}\label{subsec:staged}
In an operating plant, an end-use pressure below the requirement stops production, so closed-loop tests of an unvalidated controller are not acceptable. Following \citet{Lassahn2026}, who for the same reason verified compressor control for a large manufacturing plant with real-world data in a validated simulation environment, the method is assessed in three stages\footnote{Apart from the pipe-network simulator, the simulation, analysis and plotting code of this paper was written with Claude Opus 5.5 (Anthropic) from the authors' algorithms and checked by the authors.} before closed-loop trials in the plant, the final stage (iv): (i) identification and prediction accuracy on plant operating data; (ii) closed-loop evaluation on a pipe-network simulator that is compared with measurements from the same plant and driven by the demand time series reconstructed at the controlled end-use point; and (iii) implementation and execution-time measurement on the actual sequencing controller. Section~\ref{sec:discussion} discusses what each stage establishes and what only plant trials can confirm.

\section{Compressed air system and problem statement}\label{sec:system}

\subsection{System configuration and conventional operation}\label{subsec:system_overview}
\begin{figure*}[pos=!tp]
 \centering
 \includegraphics[width=0.9\linewidth]{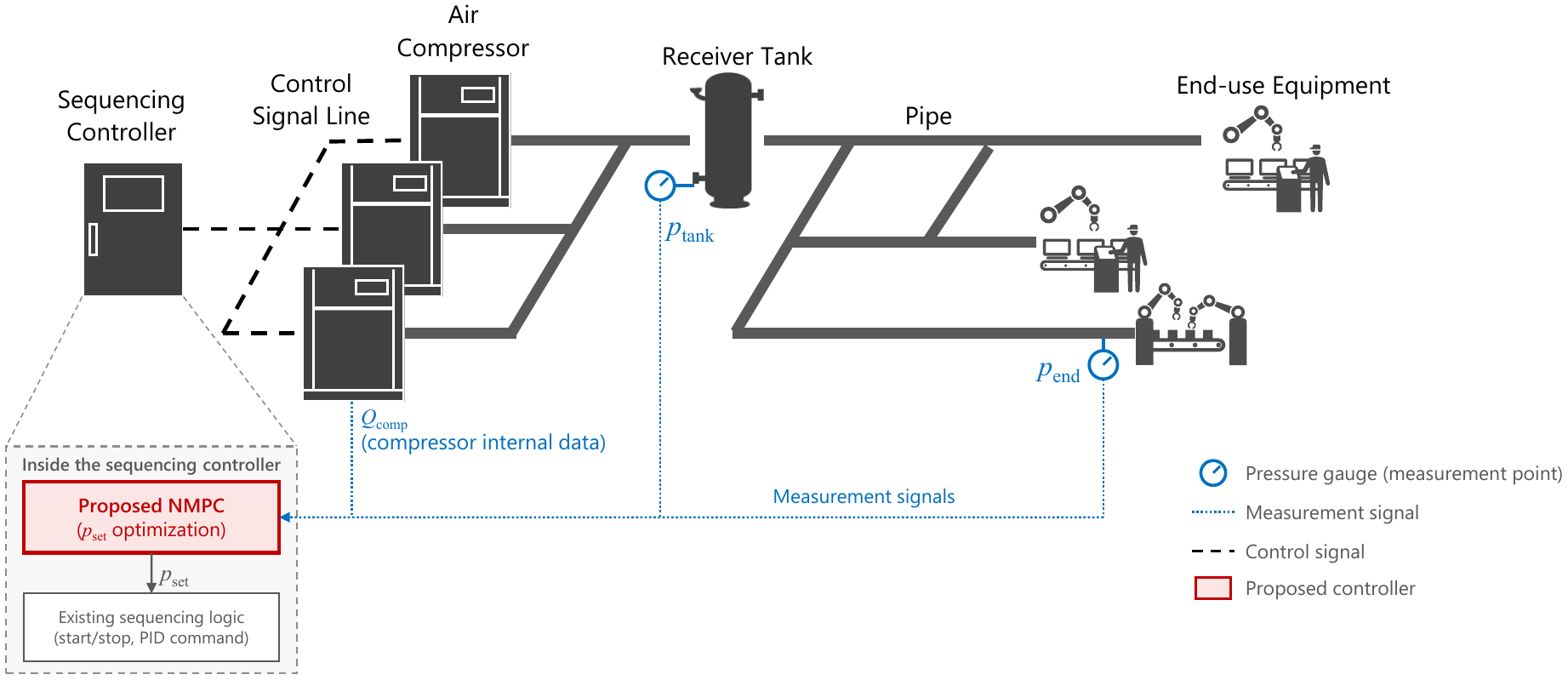}
 \caption{Configuration of the compressed air system and measurement points. The receiver pressure $\ptank$ and the end-use pressure $\pend$ are measured with pressure gauges, and the discharge flow $\Qcomp$ is obtained as compressor internal data (blue dotted lines). The proposed NMPC runs inside the existing sequencing controller (enlarged, lower left) and supplies $\pset$ to the existing sequencing logic. Adapted from \citet{Watanabe2026rengo} and drawn with Anthropic's Claude Opus 5.5.}
 \label{fig:compressed_air_system}
\end{figure*}

Fig.~\ref{fig:compressed_air_system} shows the configuration considered in this study. Several compressors coordinated by a multi-compressor sequencing controller discharge into a receiver tank, from which a pipe network distributes compressed air to end-use equipment such as production machines and air blow nozzles. The sequencing controller sends start/stop commands and PID commands for the discharge flow to each compressor. Each piece of end-use equipment needs a minimum supply pressure; if it is not met, the equipment stops and production may halt. The only variable manipulated in this study is the set pressure $\pset$, a single parameter of the sequencing controller. The control loops inside the compressors and the sequencing logic are left unchanged.

In conventional operation the set pressure is fixed well above the required pressure (Section~\ref{subsec:background}); in the target operation it is lowered dynamically as far as the required end-use pressure allows (Fig.~S1 of the Supplementary material).

\subsection{Target plant and available measurements}\label{subsec:plant}
\begin{figure*}[pos=!tp]
 \centering
 \includegraphics[width=\linewidth]{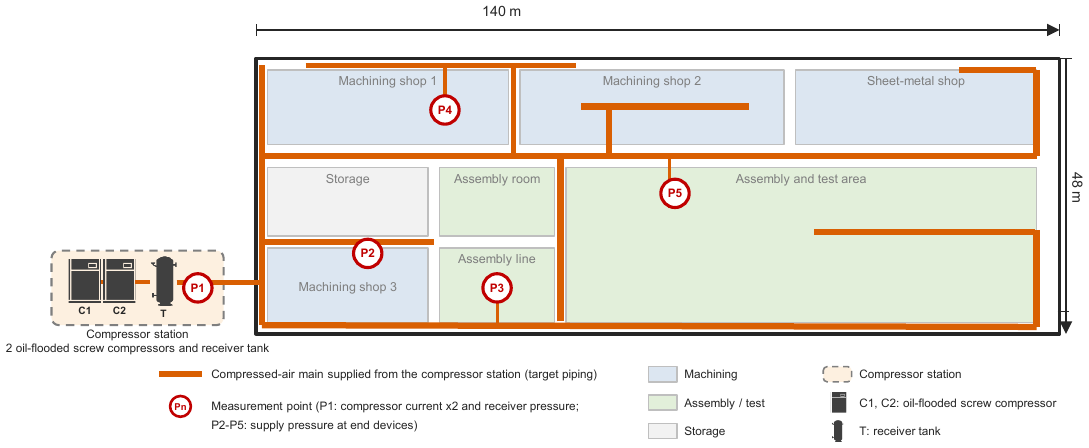}
 \caption{Plant configuration and measurement points. Orange lines: compressed-air main from the compressor station; P1: compressor currents and receiver pressure; P2--P5: end-use supply pressures. Drawn with Anthropic's Claude Opus 5.5.}
 \label{fig:plant}
\end{figure*}

Fig.~\ref{fig:plant} shows the plant used for validation, a machining and assembly factory of about 140~m $\times$ 48~m. It has two independent compressed-air systems of the same configuration; the figure shows the one modeled in the simulator of Section~\ref{subsec:setup}. Two oil-flooded screw compressors (C1, C2), one with variable speed and one with fixed speed in load/unload operation, discharge into a common receiver tank (T), and the main piping supplies end-use equipment in the machining shops, the sheet-metal shop and the assembly and test area. Five points are measured: P1 in the compressor station (the currents of both compressors and the receiver pressure) and the supply pressures P2--P5 of end-use equipment. Pressures were logged as gauge voltages (HIOKI LR5042; resolution 0.25~kPa, accuracy $\pm 0.5$\,\% of reading) and currents with clamp loggers (LR5051), both at 1~s.
 The pressure at P5, the supply point of an impact wrench in the assembly and test area, is used as the end-use pressure $\pend$; over the eight days it showed the deepest dips of the measured end-use points (minimum 0.515~MPa), although P3 was often slightly lower.

Table~\ref{tab:signals} lists the available signals. The receiver pressure $\ptank$ and the end-use pressure $\pend$ at one representative critical end-use point are measured with pressure gauges. The discharge flow $\Qcomp$ is converted from the compressor currents read over the network (rated flow per rated current, saturating at the rated flow); this common estimate of the load factor was not compared with a flow meter.
 The set pressure $\pset$ is the output of the sequencing controller itself and needs no measurement. All signals are sampled at 1~s. The only added instrumentation is therefore one pressure gauge at the end-use point.

\begin{table}[!tbp]
\caption{Available signals.}
\label{tab:signals}
\centering
\footnotesize
\begin{tabular}{@{}p{1.8cm}lp{3.0cm}p{1.2cm}@{}}
\toprule
Signal & Symbol & Source & Added cost \\
\midrule
Receiver pressure & $\ptank$ & Existing gauge (input of the sequencing controller) & None \\
End-use pressure & $\pend$ & One gauge added at the end-use point & One gauge \\
Discharge flow & $\Qcomp$ & Compressor currents (network) & None \\
Set pressure & $\pset$ & Sequencing controller (own output) & None \\
\bottomrule
\end{tabular}
\end{table}

\subsection{Problem statement}\label{subsec:problem}
The problem is to keep the set pressure $\pset(t)$ as low as possible without letting the end-use pressure fall below the required pressure $\pneed$; it is formulated in Section~\ref{sec:control}. The four practical constraints of Section~\ref{subsec:background} are adopted as design requirements.
\begin{description}
\item[R1 (no piping information)] The diameters, lengths and connections of the pipes are not used.
\item[R2 (no flow meters)] No flow meter is installed at end-use points; only the signals in Table~\ref{tab:signals} are used.
\item[R3 (existing controller)] The control law runs on the existing sequencing controller and updates the set pressure with a period well below the compressor response time of about 6~s (the 6.15-s acceleration time set in the inverter).

\item[R4 (facility changes)] No manual retuning of the controller is required after changes to the piping or the receiver.
\end{description}
The model structure of Section~\ref{sec:model} addresses R1 and R2, the controller structure of Section~\ref{sec:control} addresses R3, and the model structure addresses R4 (Section~\ref{subsec:facility_change}); re-identification (Section~\ref{subsec:reidentification}) covers changes in the dynamics themselves.

\section{Physics-constrained data-driven modeling}\label{sec:model}

\subsection{Modeling approach}\label{subsec:model_policy}
To predict the future end-use pressure under R1 and R2, no physical model of the pipe network is built. Instead, three low-order models are chained (Fig.~S2 of the Supplementary material shows their signal connections).
 The compressor model predicts the discharge flow $\Qcomp$ from $\pset$ and $\ptank$, the receiver model predicts $\ptank$ from $\Qcomp$ and the total demand $\Qtank$, and the end-use model predicts $\pend$ from $\ptank$. The demand $\Qtank$ is an exogenous input estimated from measurements (Section~\ref{subsec:demand}). The three models have six states in total (Section~\ref{subsec:computation}) and form the prediction chain of the controller directly.

\subsection{Receiver pressure model}\label{subsec:tank_model}
Assuming mass conservation and isothermal conditions, the receiver pressure follows the difference between the inflow from the compressors and the outflow to the plant, that is, the total demand \citep{fluid_power_control}:
\begin{equation}
 \frac{d\ptank(t)}{dt} = \frac{p_{\mathrm{atm}}}{\Vtank}\left(\Qcomp(t)-\Qtank(t)\right)
 \label{eq:tank_dynamics}
\end{equation}
where $\Vtank$ is the receiver volume [L], $\Qcomp(t)$ the discharge flow [L/s], $\Qtank(t)$ the outflow [L/s] and $p_{\mathrm{atm}}=0.1013$~MPa the atmospheric pressure. Equipment ratings and load conditions are given in L/min, as is customary, and converted to L/s in the equations. $\Qcomp$ is taken from the compressor internal data (Table~\ref{tab:signals}).

\subsection{Compressor discharge flow model}\label{subsec:comp_model}
A variable-speed compressor (1) regulates its output by feedback on the difference between $\pset$ and the current pressure, (2) is limited by its rated maximum flow $\Qspec$, and (3) delivers the outflow at steady state. These properties are represented by the following nonlinear first-order model with a sigmoid switching function:
\begin{align}
 \frac{d\Qcomp}{dt} &= \alpha\left(-\Qcomp(t)+\Qspec\right) f(\Delta p) \nonumber\\
 &\quad + \alpha\left(1-f(\Delta p)\right)\left(-\Qcomp(t)+\Qtank(t)\right) \label{eq:comp_dynamics}\\
 f(\Delta p) &= \frac{2}{1+\exp\left(-\beta(\pset(t)-\ptank(t)+\gamma)\right)}-1 \label{eq:sigmoid}
\end{align}
The identified parameters are $\alpha$, $\beta$ and $\gamma$: $1/\alpha$ is the time constant of the flow response, $\beta$ the sharpness of saturation with respect to the pressure difference, and $\gamma$ a pressure offset. They minimize the one-step prediction error at the sampling time $\Delta t=1$~s:
\begin{align}
 \min_{\alpha,\beta,\gamma}\ &\sum_{i=1}^{N_{\mathrm{data}}}\left(Q_{\mathrm{comp,data}}(t_i+\Delta t)-\Qcomp(t_i+\Delta t)\right)^2 \label{eq:comp_optimization}\\
 \text{s.t.}\ &\alpha>0,\quad \beta\ge 29.44 \nonumber
\end{align}
The lower bound on $\beta$ reflects the observed behavior that the compressor runs practically at full capacity when the set pressure exceeds the receiver pressure by 0.1~MPa. Because Eq.~\eqref{eq:sigmoid} equals $\tanh(\beta\Delta p/2)$, requiring at least 90\,\% saturation at $\Delta p=0.1$~MPa gives $\beta\ge 2\tanh^{-1}(0.9)/0.1=29.44$.

Eq.~\eqref{eq:comp_dynamics} thus represents the receiver-pressure PI and the compressors of Fig.~\ref{fig:control_architecture} as one identified block from $\pset$ and $\ptank$ to $\Qcomp$: it describes their joint response rather than the PI law, so the PI action is contained in $\alpha$, $\beta$ and $\gamma$. With a known PI modeled explicitly instead, the mean end-use pressure changed by 0.001~MPa (details in Section~S3 of the Supplementary material).

\subsection{End-use pressure model: second-order ARX with a pressure-drop variable}\label{subsec:end_model}
To satisfy R1, the relation between $\ptank$ and $\pend$ is represented by a data-driven ARX model \citep{Ljung1999,Soderstrom1989}. Pressure transients in a pipe propagate as waves at the speed of sound because of the compressibility and inertia of the air, and they are described by second-order differential equations \citep{fluid_power_control,Fluid_transients_in_systems}. The model order is therefore set to two.

Besides the receiver pressure itself, an auxiliary input $\pdiff$ that emphasizes changes in the pressure difference (pressure drop) between the receiver and the end-use point is introduced:
\begin{align}
\boldsymbol{x}_{\mathrm{end}}(t+\Delta t) &= A_{\mathrm{end}}\,\boldsymbol{x}_{\mathrm{end}}(t)+B_{\mathrm{end}}\begin{bmatrix}\ptank(t)\\ \pdiff(t)\end{bmatrix}
\label{eq:arx_model}\\
A_{\mathrm{end}} &= \left[\begin{smallmatrix} a_{\mathrm{end}1} & a_{\mathrm{end}2}\\ 1 & 0\end{smallmatrix}\right],\quad
B_{\mathrm{end}} = \left[\begin{smallmatrix} b_{\mathrm{end}1} & b_{\mathrm{end}2}\\ 0 & 0\end{smallmatrix}\right] \nonumber\\
\pdiff(t) &:= 2\left(\ptank(t)-\pend(t)\right) \nonumber\\
&\qquad -\left(\ptank(t-\Delta t)-\pend(t-\Delta t)\right)
\label{eq:pdiff}
\end{align}
where $\boldsymbol{x}_{\mathrm{end}}(t)=[\pend(t),\ \pend(t-\Delta t)]^\top$. Eq.~\eqref{eq:pdiff} is the pressure drop $\Delta p_{\mathrm{loss}}=\ptank-\pend$ plus its first difference, and it equals the pressure drop itself at steady state. Eq.~\eqref{eq:arx_model} is thus a model of the pressure-drop dynamics: extending pipes or changing their diameters changes the level of the pressure drop, not this structure, which is the basis of contribution~3. By the steady-state equalities (Section~\ref{subsec:constraints}), $\pend=\ptank-\pdiff$ at steady state irrespective of the parameters, so a change in the pressure-drop level is carried by the measured $\pdiff$; the parameters describe only the transient (delay and overshoot).

Because $\pdiff$ is a linear combination of past values of $\ptank$ and $\pend$, the unconstrained regressor space is identical to that of a standard ARX(2,2) model. Introducing $\pdiff$ therefore adds no information; it is a reparameterization that makes the pressure drop an explicit variable. What distinguishes the proposed model is the set of constraints in the next section and their physical interpretation.

\subsection{Physical constraints and their meaning}\label{subsec:constraints}
Plant data contain noise, so an ARX model identified only by minimizing the prediction error can produce unphysical oscillation or divergence in multi-step prediction. In real pipes, a sudden increase in air consumption at an end-use point causes a transient with overshoot, mainly because of the inertia of the air, after which the pressure difference between the receiver and the end-use point converges to a constant \citep{fluid_power_control,Fluid_transients_in_systems}. The parameters are therefore identified by the following constrained least-squares problem, which enforces this response shape and stability in multi-step prediction \citep{Nocedal2006}:
\begin{align}
 \min_{a,b}\ &\sum_{i=1}^{N_{\mathrm{data}}}\left(p_{\mathrm{end,data}}(t_i+\Delta t)-p_{\mathrm{end,pred}}(t_i+\Delta t)\right)^2 \label{eq:end_optimization}\\
 \text{s.t.}\ & a_{\mathrm{end}1}^2+4a_{\mathrm{end}2}<0 \label{eq:c_overshoot}\\
 & \left|\frac{a_{\mathrm{end}1}\pm\sqrt{a_{\mathrm{end}1}^2+4a_{\mathrm{end}2}}}{2}\right|<1 \label{eq:c_stability}\\
 & b_{\mathrm{end}1}+b_{\mathrm{end}2}=0 \label{eq:c_steady_b}\\
 & 1-a_{\mathrm{end}1}-a_{\mathrm{end}2}+b_{\mathrm{end}2}=0 \label{eq:c_steady_a}\\
 & a_{\mathrm{end}1}<0,\quad a_{\mathrm{end}2}<0 \label{eq:c_sign}
\end{align}
Table~\ref{tab:constraints} summarizes the physical meaning of each constraint and the symptom that appears when it is violated (admissible region in Fig.~S3 of the Supplementary material). Eqs.~\eqref{eq:c_steady_b} and \eqref{eq:c_steady_a} ensure that the model has an equilibrium with a constant pressure drop at steady state ($\ptank(t)=\ptank(t-\Delta t)$ and $\pend(t+\Delta t)=\pend(t)=\pend(t-\Delta t)$, so that $\pdiff=\ptank-\pend$). Under \eqref{eq:c_overshoot} the poles are complex with $|\lambda|^2=-a_{\mathrm{end}2}$, so \eqref{eq:c_stability} is equivalent to $a_{\mathrm{end}2}>-1$ and the sign condition on $a_{\mathrm{end}2}$ is implied; the general forms, which Algorithm~1 imposes, state the meaning of each constraint. The stability condition makes the prediction stable in the way the NMPC uses the model, with $\pdiff$ held at its measured value (Section~\ref{subsec:mpc_formulation}); the model is not meant to predict the pressure drop itself, and computing $\pdiff$ from its output would place the poles at 1 and 1.047 (Table~\ref{tab:identified}). Between the receiver and a nearly closed end-use point, a pressure disturbance travels at the speed of sound (about 340~m/s in air, depending on temperature only) and the fundamental period of the line is $4L/c$ \citep{Streeter1985}; the condition $a_{\mathrm{end}1}<0$ (period below $4\Delta t$) therefore holds for pipe lengths below $c\Delta t\approx 340$~m, which covers a centralized station at the scale of a factory building (about 140~m $\times$ 48~m here). Only this order of magnitude is used, not piping data (R1).

\begin{table}[!tbp]
\caption{Physical constraints imposed on the end-use pressure model.}
\label{tab:constraints}
\centering
\footnotesize
\setlength{\tabcolsep}{3pt}
\begin{tabular}{@{}p{1.9cm}cp{2.5cm}p{2.5cm}@{}}
\toprule
Constraint & Eq. & Physical meaning & Symptom if violated \\
\midrule
Overshoot condition & \eqref{eq:c_overshoot} & Complex-conjugate poles of the AR part (damped oscillation) & Monotone response; the pressure drop right after a load step is underestimated \\
Stability & \eqref{eq:c_stability} & AR poles inside the unit circle & Divergent multi-step prediction (Section~\ref{subsec:multistep}) \\
Steady-state equality (input) & \eqref{eq:c_steady_b} & At steady state, $\ptank$ and $\pdiff$ act only through the pressure drop & Unphysical offset proportional to the receiver pressure \\
Steady-state equality (equilibrium) & \eqref{eq:c_steady_a} & An equilibrium with constant pressure drop exists & Drift at steady state \\
Sign condition ($a_{\mathrm{end}2}$) & \eqref{eq:c_sign} & Sign of the inertia term (restoring force); implied by \eqref{eq:c_overshoot} & --- \\
Sign condition ($a_{\mathrm{end}1}$) & \eqref{eq:c_sign} & Oscillation period below $4\Delta t$, as for pressure waves in pipes shorter than $c\Delta t\approx 340$~m ($4L/c<4$~s) & Period of $4\Delta t$ or longer; the fit moves to the double-root boundary \\

\bottomrule
\end{tabular}
\end{table}

\subsection{Identification procedure}\label{subsec:identification}
Algorithm~1 summarizes the identification of the three models. Apart from the margin of $10^{-8}$ for the strict inequalities, the lower bound on $\beta$ and the choice of the data windows, it has no tuning parameters; for the same data and initial guess it returns the same model, in about 0.03~s for the end-use model of Section~\ref{subsec:onestep}.
 The re-identification of Section~\ref{subsec:reidentification} simply reruns this procedure.

\begin{table}[!tbp]
\small
\noindent\textbf{Algorithm 1}\ Physics-constrained identification\par\vspace{0.5mm}\hrule\vspace{1mm}
\begin{enumerate}[1.]
\item \textbf{Data}: 1-s samples of $\ptank$, $\pend$, $\Qcomp$ and $\pset$. The compressor model uses an interval that contains a set-pressure change; the end-use model uses normal operation that contains demand (pressure-drop) variation.
\item \textbf{Preprocessing}: none. Measurements are used as recorded, without gap filling, outlier removal or filtering.
\item \textbf{Receiver model}: $\Vtank$ is a known rating; no parameter is identified.
\item \textbf{Compressor model}: solve Eq.~\eqref{eq:comp_optimization} by sequential quadratic programming (SQP).
\item \textbf{End-use model}: solve Eqs.~\eqref{eq:end_optimization}--\eqref{eq:c_sign} by SQP (MATLAB \texttt{fmincon}, default tolerances, initial guess $(a_{\mathrm{end}1},a_{\mathrm{end}2},b_{\mathrm{end}1},b_{\mathrm{end}2})=(-0.5,-0.2,0.1,-0.1)$), with two linear equality constraints and the inequality constraints; the sign conditions are imposed as $\le -10^{-8}$.

\item \textbf{Check}: record constraint satisfaction (pole magnitudes, overshoot condition) and the one-step mean absolute error (MAE) in the identification and validation windows.
\end{enumerate}
\vspace{1mm}\hrule
\end{table}

\section{Lightweight hierarchical nonlinear MPC}\label{sec:control}

\subsection{Control architecture}\label{subsec:architecture}
\begin{figure*}[pos=!tp]
 \centering
 \includegraphics[width=\linewidth]{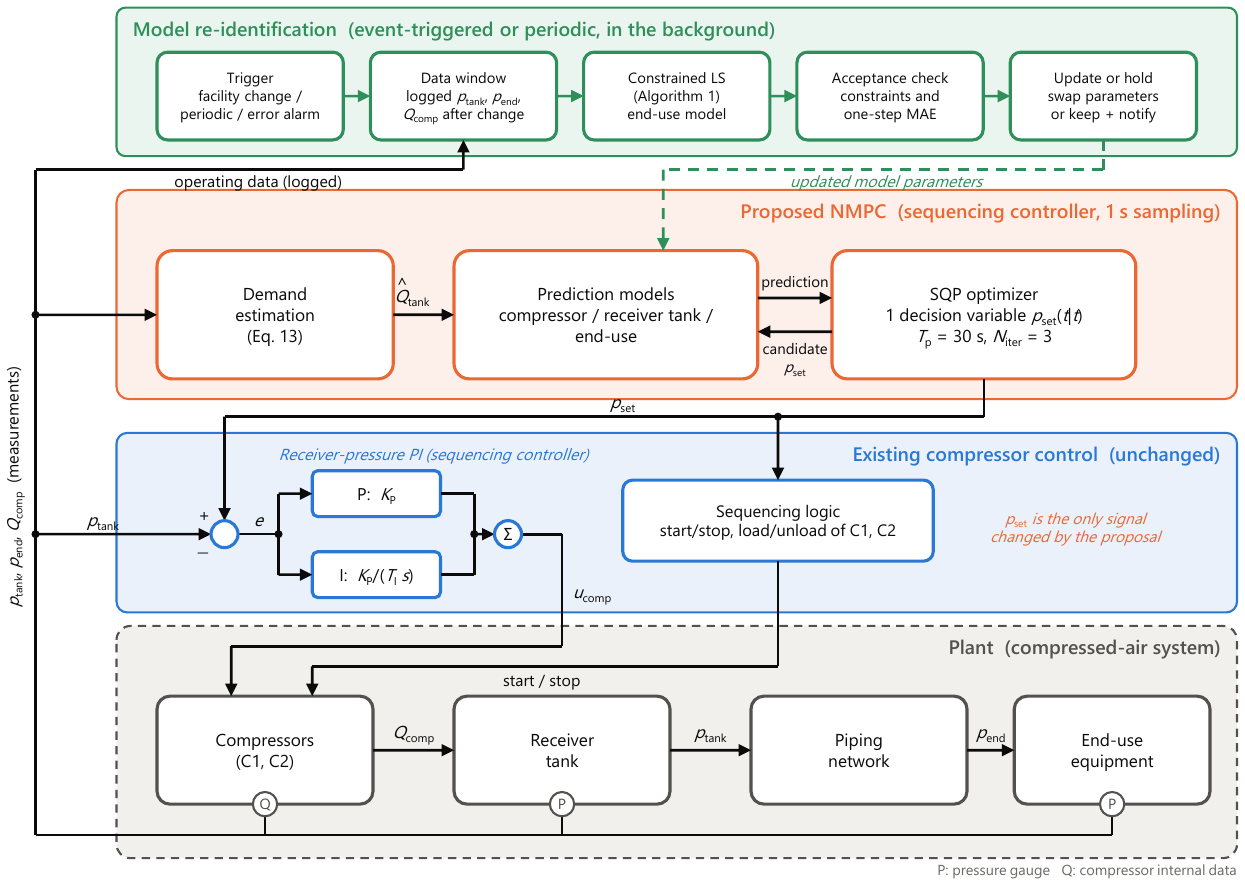}
 \caption{Proposed control architecture. The supervisory NMPC (second row) runs on the existing sequencing controller and determines $\pset$ from the 1-s measurements ($\ptank$, $\pend$, $\Qcomp$) by demand estimation, three-model prediction and SQP. The existing compressor control (third row) and the plant (bottom) are unchanged. Re-identification (top, Section~\ref{subsec:reidentification}) reruns Algorithm~1 on logged data when triggered and replaces the parameters if the check is passed. Drawn with Anthropic's Claude Opus 5.5.}
 \label{fig:control_architecture}
\end{figure*}

The proposed controller (Fig.~\ref{fig:control_architecture}) is a lightweight nonlinear MPC in the upper layer that optimizes the set pressure $\pset$ online and is intended for implementation in the existing sequencing controller. The lower layer consists of the existing sequencing logic and the receiver-pressure PI control, which make the receiver pressure track the $\pset$ given by the upper layer. This hierarchy, in which an upper layer predicts the disturbance (here, the pressure drop to the end-use point) and adjusts the references of local controllers, is widely used in supervisory MPC and real-time optimization in the process industries \citep{Garcia1989,plant_MPC1,plant_MPC2,Darby2012}. ``Lightweight'' means that the design is kept to the minimum needed for implementation: three kinds of measured signals, six model states and a single decision variable in the online optimization (R3).

\subsection{Demand estimation}\label{subsec:demand}
The controller estimates the total demand of the plant from the current discharge flow $\Qcomp$ and receiver pressure $\ptank$ by inverting Eq.~\eqref{eq:tank_dynamics}:
\begin{equation}
 \hat{Q}_{\mathrm{tank}}(k)=\Qcomp(k-1)-\frac{\Vtank}{p_{\mathrm{atm}}}\,\frac{\ptank(k)-\ptank(k-1)}{\Delta t}
 \label{eq:demand_estimate}
\end{equation}
The difference is a 1-s backward difference, and no moving average or other filter is applied. The demand is assumed constant over the prediction horizon. Demand changes slowly relative to the horizon $T_p=30$~s, and after a sudden change the estimate is updated in the next control cycle, so the effect of this assumption is limited (see Section~\ref{subsec:limitations} for its limits).

\subsection{NMPC formulation}\label{subsec:mpc_formulation}
At each control cycle, the optimizer computes the set pressure $\pset(t)$ so that the predicted end-use pressure tracks the target pressure $\pref$, defined as the required pressure $\pneed$ plus a margin. It minimizes the squared deviation of the predicted end-use pressure from the target over the horizon $T_p$. The control horizon is $T_c=1$, so the set pressure is held constant over the prediction horizon, $\pset(t+k|t)=\pset(t|t)$ for $k=1,\dots,T_p-1$:
\begin{align}
 \min_{\pset(t|t)}\ J &= J_{\mathrm{stage}}+J_{\mathrm{terminal}} \label{eq:mpc_objective}\\
 J_{\mathrm{stage}} &= \sum_{k=1}^{T_p-1} w_{\mathrm{err}}\left(\pend(t+k|t)-\pref\right)^2 \label{eq:mpc_stage}\\
 J_{\mathrm{terminal}} &= T_p\, w_{\mathrm{term}}\left(\pend(t+T_p|t)-\pref\right)^2 \label{eq:mpc_terminal}\\
 \text{s.t.}\ & p_{\mathrm{set,min}}\le\pset(t|t)\le p_{\mathrm{set,max}} \label{eq:mpc_constraint_u}
\end{align}
The decision variable is the scalar $\pset(t|t)$, which is recomputed at every cycle and applied until the next one (receding horizon). $\pend(t+k|t)$ is the end-use pressure predicted $k$ steps ahead by the chain of the three models; the pressure-drop variable $\pdiff$ of Section~\ref{subsec:end_model} is computed from measurements at the start of the horizon and held constant over it (pressure-drop hold). Because the pressure drop at a given flow grows as the absolute pressure falls (by about 1.08\,\% per 1\,\% in the simulator, Section~\ref{subsec:real_demand}), a set-pressure change of 0.02~MPa within the horizon changes the held drop by about 3\,\%, or about 1~kPa at 35~kPa, which is small compared with the margin $\pref-\pneed$.

The weights have physical interpretations. The ratio $w_{\mathrm{term}}/w_{\mathrm{err}}=10$ expresses how much more the steady-state deviation at the end of the horizon matters than the stage deviations; with the factor $T_p$, the terminal term (weight 300) dominates the stage terms (29), so the set pressure is chosen mainly to bring the end-use pressure to $\pref$ at the end of the horizon. Holding the decision variable over the horizon is a move-blocking strategy \citep{Cagienard2007}, and $T_p=30$~s covers the dead time of about 20~s of the end-use pressure response (Appendix~\ref{app:pid}).
 The set pressure is lowered by tracking the low target $\pref$, not by a penalty on itself (the implemented controller adds $T_p w_{\mathrm{u}}\pset^2$ with $w_{\mathrm{u}}=10^{-11}$, about $1.5\times10^{-10}$, which has no effect).
 The end-use lower bound is not an explicit state constraint, and power does not appear in the objective; the required pressure is ensured by the margin in $\pref$, so the only constraints are the set-pressure bounds, which keeps the computation light (Section~\ref{subsec:limitations}).

\subsection{Computation and implementation}\label{subsec:computation}
Problem~\eqref{eq:mpc_objective}--\eqref{eq:mpc_constraint_u} is solved online at every control cycle by SQP. Two measures make this feasible on the limited computing resources of the sequencing controller (Algorithm~S1 of the Supplementary material).

First, gradients are computed efficiently. Each SQP iteration needs the Jacobians with respect to the states and the input of the chained state equation $\dot{\boldsymbol{x}}=F(\boldsymbol{x},\pset)$, with states $\boldsymbol{x}=[\Qcomp,\ptank,\pend(k),\pend(k-1),\hat{Q}_{\mathrm{tank}},\pdiff]^\top$. Analytic Jacobians of all terms, including the sigmoid function~\eqref{eq:sigmoid}, are supplied to the solver, so no model evaluations for numerical differentiation are needed.

Second, the initial guess and the number of iterations are controlled. The optimal set pressure of the previous cycle is used as a warm start. At start-up, or when the solver has not found a feasible solution, the initial guess is
\begin{equation}
 \pset^{(0)}(t)=\ptank(t)-\left(\pend(t)-\pref\right)
 \label{eq:initial_guess}
\end{equation}
which is the set pressure that brings the end-use pressure to $\pref$ if the current pressure drop persists, and is therefore close to the optimum. The number of SQP iterations is capped at $N_{\mathrm{iter}}=3$; when the cap is reached, the current iterate is applied. Because the only constraints are the bounds on the set pressure, every iterate is feasible. The MATLAB implementation used in the simulations also treats the predicted states as variables and keeps the previous set pressure when the model equations are not met within the tolerance (Section~\ref{subsec:artificial}).
 This follows the idea of the real-time iteration scheme, which applies solutions obtained with a fixed number of iterations \citep{Diehl2005}. Each iteration costs one $T_p$-step forward simulation of the three models and one evaluation of the analytic Jacobians, so the computational load is $O(T_p N_{\mathrm{iter}})$. MPC on hardware with limited computing resources has been pursued through fast online optimization \citep{Wang2010} and implementation on low-level control hardware \citep{Huyck2014}; this study achieves it through a single decision variable and analytic Jacobians.

\subsection{Coping with facility changes: structure and re-identification}\label{subsec:reidentification}
The answer of the proposed method to R4 is robustness through the model structure, complemented by an operating procedure for re-identification. Because the model of Section~\ref{subsec:end_model} describes the pressure-drop dynamics, the prediction of the initial model does not break down structurally after pipes are extended, diameters are changed or a receiver is added; a change in the pressure-drop level is absorbed by the online estimate of Eq.~\eqref{eq:pdiff} (Section~\ref{subsec:facility_change}). In a real plant, however, changing pipe routes or diameters can alter the pressure-drop dynamics themselves (delay and overshoot). Such changes appear in the parameters, so the model is updated by rerunning Algorithm~1 on recent operating data with the following procedure.
\begin{enumerate}[(i)]
\item Trigger: a facility-change event (completion of expansion work), a periodic schedule (e.g., monthly), or the one-step prediction error exceeding a threshold.
\item Data window: operating data accumulated after the change. Its length is not fixed; it must contain enough demand (pressure-drop) variation, as required by Algorithm~1. Right after the change, data can be accumulated while the existing constant set-pressure operation continues.
\item Identification: rerun the end-use model identification of Algorithm~1. The receiver volume is updated directly as a known rating; the compressor model is unchanged.
\item Check: constraint satisfaction, an improvement of the one-step MAE in the window and whether both $a_{\mathrm{end}1}$ and $a_{\mathrm{end}2}$ lie on their sign bounds, a sign of too little excitation, are checked automatically.
\item Replacement: if the check is passed, the prediction model is updated; otherwise the update is withheld and the operator is notified.
\end{enumerate}
Section~\ref{subsec:facility_change} evaluates the NMPC with its initial model under plant expansion and network split.

\section{Validation on plant data}\label{sec:validation}

\subsection{Dataset}\label{subsec:dataset}
The operating data of this section were acquired in the other compressed-air system of the plant (Section~\ref{subsec:plant}), not in the one modeled by the simulator: because the piping differs from plant to plant, the identification procedure is validated on a system other than the one used for the closed-loop evaluation, and Section~\ref{sec:closedloop} applies the same procedure to the simulator configuration. They cover
 eight consecutive days sampled at 1~s, including the receiver pressure, the end-use pressure and the operating signals of the compressors. The end-use model is identified on about 9.5~h from midnight to 09:28 on a production day (34,113 samples) and validated on the daytime interval from 10:40 to 16:07 of the same day (19,623 samples).
 The compressor model is identified on about 4 minutes (258 samples) around a change of the set pressure of the variable-speed compressor from 0.60 to 0.65~MPa, made for this purpose: $\alpha$ and $\beta$ need set-pressure excitation, but because the model takes the estimated demand as input, the test needs no constant load and is done once per compressor product. After the change,
 the discharge flow saturates at the rated flow $\Qspec=7300$~L/min, so the interval contains both the tracking and the saturation behavior of Eq.~\eqref{eq:comp_dynamics}. The receiver volume $\Vtank=1400$~L is used; the receiver model is determined by this known volume alone and needs no identification data.

\subsection{Identified parameters and prediction accuracy}\label{subsec:onestep}
Table~\ref{tab:identified} lists the parameters identified by Algorithm~1. The poles of the autoregressive part of the end-use model are $\pm 0.154\mathrm{i}$ (magnitude 0.154), and all constraints~\eqref{eq:c_overshoot}--\eqref{eq:c_sign} are satisfied; the active ones are the two equalities and the sign condition on $a_{\mathrm{end}1}$.

\begin{table}[!tbp]
\caption{Identified parameters. The end-use parameters are those of \citet{Watanabe2026rengo}; the compressor parameters were re-identified.}
\label{tab:identified}
\centering
\footnotesize
\begin{tabular}{llr}
\toprule
Model & Parameter & Value \\
\midrule
Compressor & $\alpha$ [1/s] & 0.2770 \\
       & $\beta$ [1/MPa] & 301.0 \\
       & $\gamma$ [MPa] & 0.00288 \\
End-use & $a_{\mathrm{end}1}$ & $-1.00\times10^{-8}$ \\
       & $a_{\mathrm{end}2}$ & $-0.02358$ \\
       & $b_{\mathrm{end}1}$ & $1.02358$ \\
       & $b_{\mathrm{end}2}$ & $-1.02358$ \\
\midrule
\multicolumn{2}{l}{Pole magnitude of the end-use model} & 0.154 \\
\multicolumn{2}{l}{$a_{\mathrm{end}1}^2+4a_{\mathrm{end}2}$} & $-0.0943<0$ \\
\bottomrule
\end{tabular}
\end{table}

\begin{figure}[pos=!tbp]
 \centering
 \includegraphics[width=\linewidth]{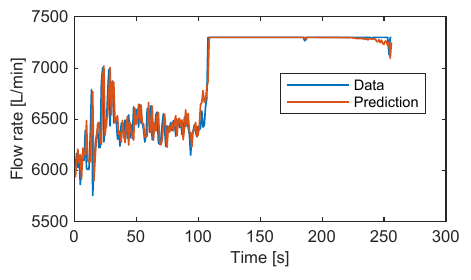}
 \caption{One-step prediction of the discharge flow by the compressor model (258~s including the set-pressure change from 0.60 to 0.65~MPa). Error metrics are given in Table~\ref{tab:onestep}.}
 \label{fig:comp_tank_model}
\end{figure}

Fig.~\ref{fig:comp_tank_model} shows the one-step (1~s ahead) predictions of the compressor and receiver models. Before the set-pressure change, the discharge flow fluctuates with the demand, and the absolute percentage error (APE) temporarily increases, up to about 18\,\%. After the change, the discharge flow saturates at the rated flow, and the model reproduces this saturation well. The receiver pressure rises after the change, and the prediction follows the measurement (Fig.~S4 of the Supplementary material).
 Over the whole interval, the mean absolute percentage error (MAPE) is 1.10\,\% for the discharge flow model and 0.03\,\% for the receiver pressure model (Table~\ref{tab:onestep}).

\begin{figure}[pos=!tbp]
 \centering
 \begin{minipage}[t]{0.48\linewidth}\centering
  \includegraphics[width=\linewidth,page=1]{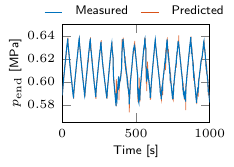}\\ {\small (a) End-use pressure}
 \end{minipage}\hfill
 \begin{minipage}[t]{0.48\linewidth}\centering
  \includegraphics[width=\linewidth,page=2]{fig05_end_model.pdf}\\ {\small (b) Pressure drop}
 \end{minipage}
 \caption{Recursive prediction with the measured pressure drop by the end-use pressure model (extracted interval with rapid pressure-drop changes, 1000~s of the identification window).}
 \label{fig:end_model}
\end{figure}

The end-use pressure model is evaluated by recursive prediction with the measured pressure drop: only $\pend$ is computed recursively from the model's own past outputs, while $\ptank$ and $\pdiff$ are taken from measurements. Because the measured pressure drop contains the measured end-use pressure, this prediction is close to a one-step prediction; it shows how well the model reproduces the transient after a measured change in the pressure drop,
 prediction with the pressure drop held, as in the NMPC, is evaluated in Section~\ref{subsec:multistep}. Over the identification window, the MAPE is 0.24\,\%, the MAE 1.43~kPa and the maximum error 25.8~kPa (one-step MAE 1.46~kPa). Fig.~\ref{fig:end_model} shows an extracted interval in which the pressure drop in the piping changes strongly as end-use equipment operates. The prediction follows the measurement even during these large transients;
 in this interval the MAPE is 0.30\,\%, the MAE 1.81~kPa and the maximum error 17.3~kPa. Fig.~\ref{fig:end_model}(b) shows that this error lies in the pressure drop (about 8~kPa on average), since the end-use pressure largely follows the 56-kPa sawtooth of the receiver pressure.

\begin{table}[!tbp]
\caption{Prediction accuracy on plant data: one-step prediction for $\Qcomp$ and $\ptank$, recursive prediction with the measured pressure drop for $\pend$ (Section~\ref{subsec:onestep}). MAE and maximum error are in the unit given for each model.}
\label{tab:onestep}
\centering
\footnotesize
\begin{tabular}{@{}lccc@{}}
\toprule
Model & MAPE [\%] & MAE & Max \\
\midrule
$\Qcomp$ [L/min] & 1.10 & 70.5 & 1017 \\
$\ptank$ [kPa] & 0.031 & 0.20 & 0.79 \\
$\pend$, whole [kPa] & 0.24 & 1.43 & 25.8 \\
$\pend$, rapid change [kPa] & 0.30 & 1.81 & 17.3 \\
\bottomrule
\end{tabular}
\end{table}

\subsection{Multi-step prediction: constrained versus unconstrained identification}\label{subsec:multistep}
In the NMPC, the measured end-use pressure enters the prediction at the 30-s horizon only through the current pressure drop, which is held over the horizon (Section~\ref{subsec:mpc_formulation}). Without it, the controller would not correct a lasting change of the drop, such as a rise in demand, and the end-use pressure would settle off its target, the problem that offset-free MPC solves by holding an estimated disturbance \citep{Muske2002,Pannocchia2003}. An end-use model is therefore usable in the NMPC only if (a) the current drop can be held in its prediction without making it unstable and (b) the held prediction is accurate. The constraints are meant to provide (a) and (b), not higher accuracy. Three models identified from the same data are compared: (M-con) the proposed model (Table~\ref{tab:identified}); (M-unc) the same model identified without the constraints (autoregressive pole 1.997); and (M-std) a standard ARX(2,2) model with the receiver pressure as its only input (poles 0.869 and 0.059). Both unconstrained models violate the steady-state equalities, and they give identical one-step predictions (MAE 0.50~kPa) because they span the same regressors.

The test uses the validation window of Section~\ref{subsec:dataset} (10:40--16:07). Every 10~s (1,951 origins), the end-use pressure is predicted 30--120~s ahead from the measured end-use pressures up to that time and the measured receiver pressure over the horizon, so that only the end-use model is tested. With hold (``Hold'' in Table~\ref{tab:multistep}), the current drop is kept constant over the horizon: for M-con and M-unc as the pressure-drop input, set at the origin to the value that reproduces the measured end-use pressure; for M-std, which has no such input, through its prediction error at the origin, added to every step. Without hold (``No hold''), possible only for M-std, the model is simply run forward.

M-con meets both requirements: held, it is stable and as accurate (1.43--1.49~kPa) as the standard model without hold (1.45~kPa). M-unc fails (a): although it fits the data as well as M-std, it diverges, because the held input feeds its unstable pole. M-std fails (b): held, its error grows to 4.5~kPa, because the error at the origin builds up through its autoregressive part to 8.1 times its size ($1/(1-a_1-a_2)$); without hold it is accurate but, after about 20~s (pole 0.869), no longer contains the measured end-use pressure and so gives the NMPC no feedback (predictions in Fig.~S5 of the Supplementary material).

M-con can be held because, with the equalities \eqref{eq:c_steady_b} and \eqref{eq:c_steady_a}, its held prediction reduces to $\hat p_{\mathrm{end}}(t+k)=a_{\mathrm{end}1}\hat p_{\mathrm{end}}(t+k-1)+a_{\mathrm{end}2}\hat p_{\mathrm{end}}(t+k-2)+(1-a_{\mathrm{end}1}-a_{\mathrm{end}2})(\hat p_{\mathrm{tank}}(t+k-1)-\pdiff(t))$, a unit-gain filter of the receiver pressure minus the held drop that settles within 2--3~s. The held input thus acts as the measured pressure drop itself, and a change in its level after a facility change is absorbed in the same way (Section~\ref{subsec:facility_change}). Least squares delivers this structure only under the constraints; what matters is the equalities and stability, not the exact poles, whose magnitudes (0.154 for the plant, below 0.06 for the simulator configuration, Section~\ref{subsec:setup}) the data determine only weakly (Table~S3).

\begin{table}[!tbp]
\caption{Multi-step prediction by identification method (1,951 origins in the validation window, MAE [kPa]). M-con: proposed constrained identification; M-unc: same regressors without constraints; M-std: standard ARX(2,2). Hold: the pressure drop at the origin (the origin residual for M-std) is held over the horizon; its 1-s value is the first step of this prediction. $|\lambda|$: magnitudes of the autoregressive poles.
}
\label{tab:multistep}
\centering
\footnotesize
\setlength{\tabcolsep}{3pt}
\begin{tabular}{@{}llcccccc@{}}
\toprule
Model & Use & $|\lambda|$ & 1~s & 30~s & 60~s & 90~s & 120~s \\
\midrule
M-con & Hold & 0.154, 0.154 & 0.49 & 1.45 & 1.49 & 1.43 & 1.45 \\
M-unc & Hold & 1.997, 0.391 & 1.31 & \multicolumn{4}{c}{Diverged} \\
M-std & No hold & 0.869, 0.059 & 0.50 & 1.45 & 1.45 & 1.45 & 1.45 \\
M-std & Hold & 0.869, 0.059 & 0.77 & 4.45 & 4.54 & 4.49 & 4.47 \\
\bottomrule
\end{tabular}
\end{table}

\subsection{Comparison of the pipe-network simulator with measurements}\label{subsec:simulator_verification}
\begin{figure}[pos=!tbp]
 \centering
 \begin{minipage}[t]{0.48\linewidth}\centering
  \includegraphics[width=\linewidth]{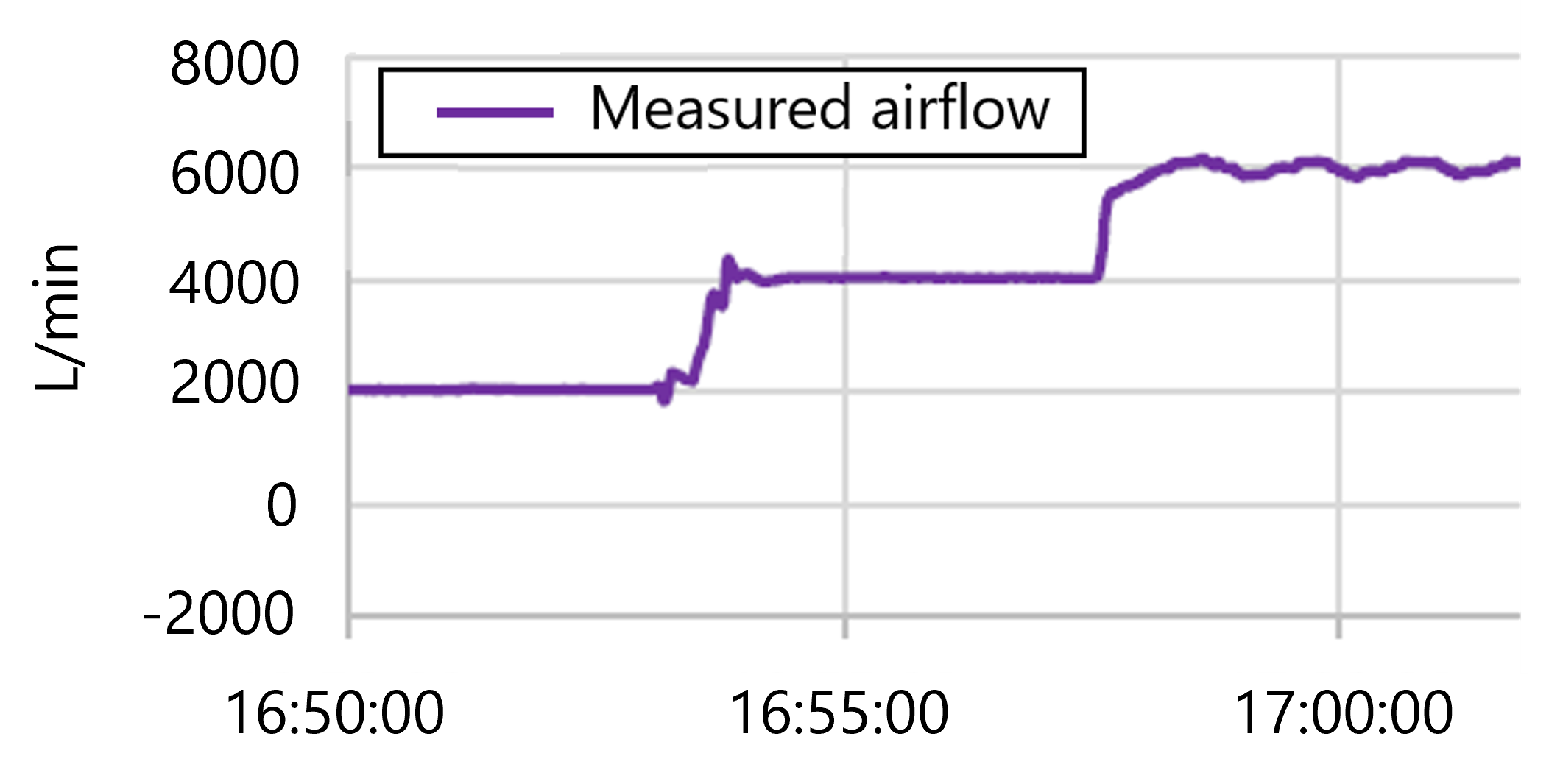}\\ {\small (a) Measured air consumption at an end-use point}
 \end{minipage}\hfill
 \begin{minipage}[t]{0.48\linewidth}\centering
  \includegraphics[width=\linewidth]{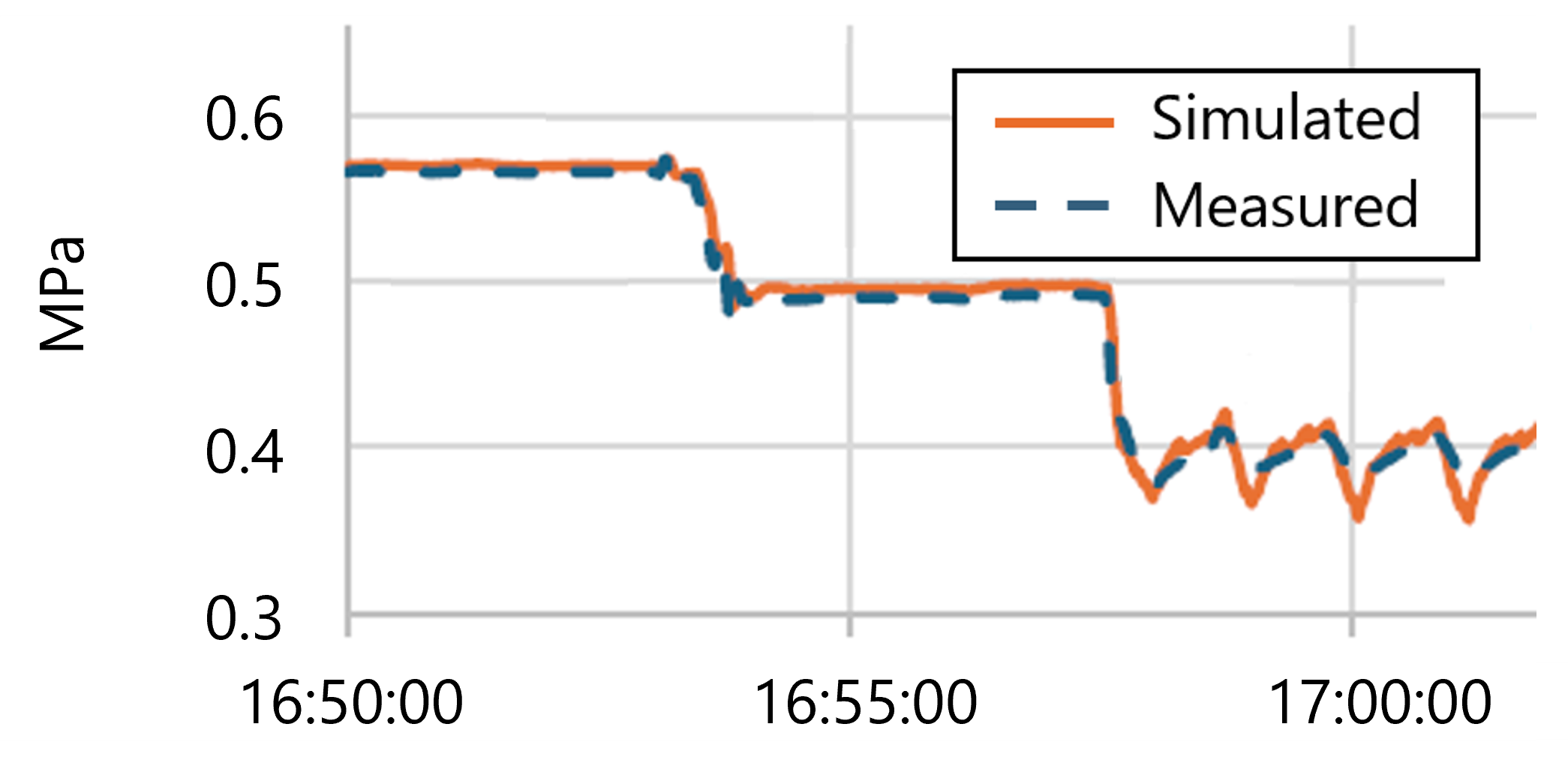}\\ {\small (b) Measured and simulated end-use pressure}
 \end{minipage}
 \caption{Comparison of measurements and simulation: (a) measured air consumption at an end-use point, used as the boundary condition; (b) measured and simulated end-use pressure (mean error 5~kPa with steady receiver pressure, 8~kPa during oscillation).}

 \label{fig:simulator_verification}
\end{figure}

The closed-loop evaluation of Section~\ref{sec:closedloop} uses a pipe-network simulator \citep{Liu2021patent} that solves unsteady one-dimensional flow with friction and heat losses in the pipes, with the conservation of mass, momentum and energy as governing equations, by the finite volume method \citep{air_system_simulator}. The receiver pressure at the inlet and the air consumption at each end-use point are the boundary conditions, and the dynamic pressure variation inside the network is computed. The simulator is used only for evaluation; the controller uses neither the piping data nor the simulator. It was built for this study from the pipe parameters of the target plant (Section~\ref{subsec:setup}); Fig.~\ref{fig:simulator_verification}, the only comparison with measurements, was made before the closed-loop evaluation, and no data were held out for a separate validation.

Fig.~\ref{fig:simulator_verification} compares measurements and simulation. When the air consumption reached 6000~L/min, the upstream compressors switched to load/unload operation, and the receiver pressure and the end-use pressure oscillated periodically. The simulation follows the measurement through these transients and oscillations; the mean pressure error is 5~kPa while the receiver pressure is steady and 8~kPa during the oscillation.

\section{Closed-loop evaluation}\label{sec:closedloop}

\subsection{Evaluation setup}\label{subsec:setup}
\begin{figure*}[pos=!tp]
 \centering
 \includegraphics[width=\linewidth]{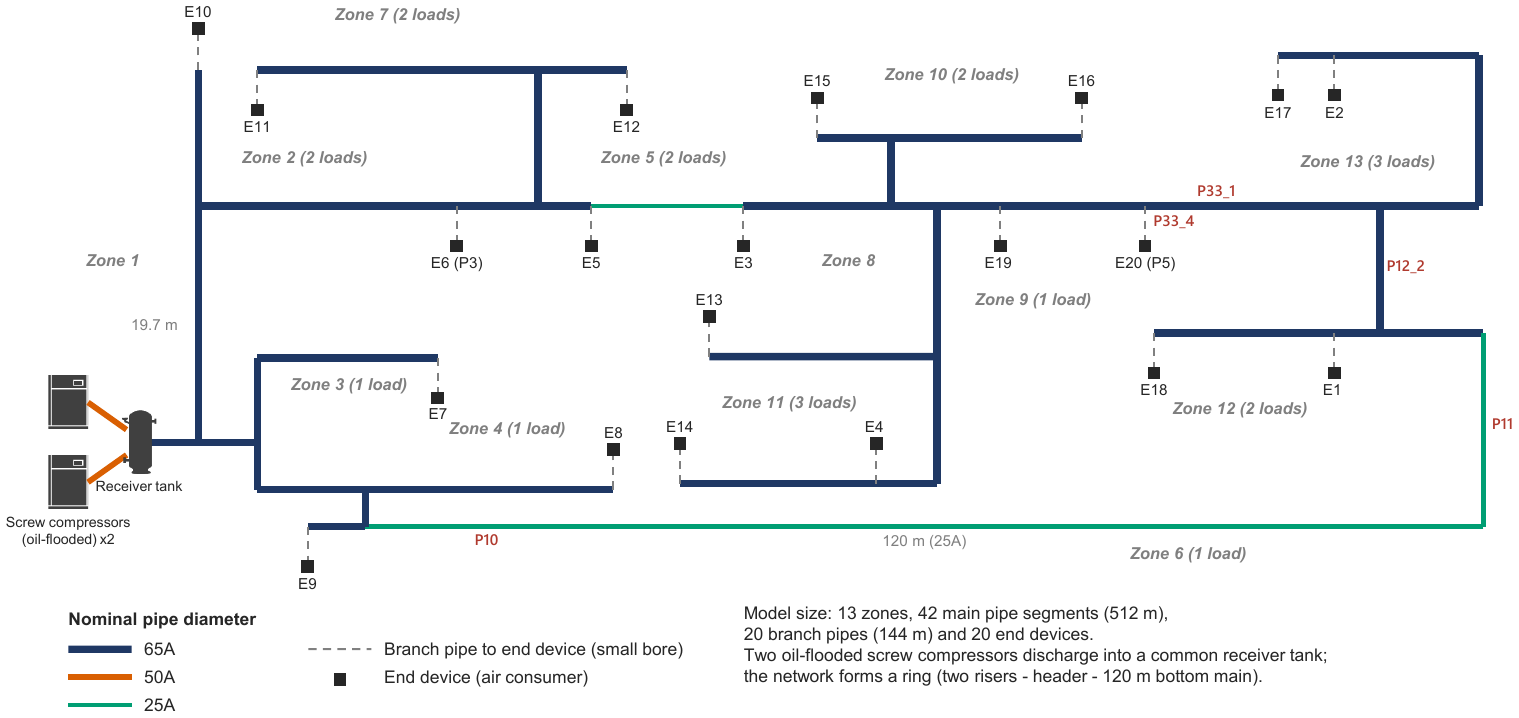}
 \caption{Pipe layout of the full simulator: 13 zones, 42 main pipe sections (512~m), 20 branch pipes (144~m) and 20 end-use points (E1--E20). Line colors indicate the nominal size (65A/50A/25A = DN65/50/25), and dashed lines are branch pipes to end-use equipment. The controlled end-use point is E20 and the neighboring end-use point is E1. Drawn with Anthropic's Claude Opus 5.5.}
 \label{fig:sim_layout}
\end{figure*}

The full simulator connects the receiver model and the compressor model of Sections~\ref{subsec:tank_model} and~\ref{subsec:comp_model} to the inlet of the pipe-network simulator of Section~\ref{subsec:simulator_verification}. It simulates the system from the compressors to the end-use points, with the set pressure and the flow consumed at each end-use point as boundary conditions. Fig.~\ref{fig:sim_layout} shows the pipe network. It models the main piping of the target plant (Fig.~\ref{fig:plant}); because the layout of the branch pipes from the main to the end-use equipment is unknown, the branch pipes were assumed so that the measured pressure drops are reproduced. The network consists of 13 zones, 42 main pipe sections (512~m in total; DN65, DN50 and DN25 with inner diameters of 67.9, 52.6 and 27.6~mm), 20 branch pipes (144~m; DN15, 16.1~mm) and 20 end-use points (62 pipes and 63 nodes). Two compressors discharge into a common receiver, and the network forms a loop through two risers, the header and a 120-m main along the bottom of the figure. The controlled end-use point is E20, which corresponds to P5 in Fig.~\ref{fig:plant}, the neighboring end-use point is E1,
 and the remaining 18 points form the base load. The cell length is 1~m and the time step 0.01~s; the simulation runs at about real time on a PC (Intel Core i7-1165G7, 2.8~GHz, 32~GB). The fixed-speed compressor adds a constant flow to that of the variable-speed one; neither a minimum speed nor unloading is modeled (Section~\ref{subsec:limitations}). Compressor flows refer to 20~$^\circ$C and end-use flows to 0~$^\circ$C and 101.325~kPa (factor 1.073). No delay or noise is added to the measurements given to the controllers.
 Each run starts with a 200-s settling period, which is excluded from the evaluation.

Table~\ref{tab:sim_settings} lists the main parameters. The required pressure $\pneed=0.50$~MPa (gauge) was set for this study, since the requirement differs between equipment; the target pressure is $\pref=0.572$~MPa (margin 0.072~MPa).
 The set pressure is limited to 0.50--0.95~MPa by the mechanical limits of the compressors. The internal end-use model of the NMPC was identified with Algorithm~1 from operating data of the simulator ($a_{\mathrm{end}1}=a_{\mathrm{end}2}=-0.0033$, $b_{\mathrm{end}1}=-b_{\mathrm{end}2}=1.0065$; only the equalities active), and its compressor and receiver models equal those of the simulator (Table~\ref{tab:sim_settings}). Its pole magnitude of 0.057 shows that the simulator produces practically no damped oscillation between the receiver and the end-use point, so the closed-loop evaluation does not include the damped oscillation of the plant model (pole magnitude 0.154, decaying within 2--3~s, short compared with $T_p=30$~s; Section~\ref{subsec:limitations}).

\begin{table}[!tbp]
\caption{Simulation and controller parameters. Pressures are gauge; the simulator and the controllers work internally with absolute pressure (atmospheric pressure 0.1013~MPa added).}
\label{tab:sim_settings}
\centering
\footnotesize
\begin{tabular}{@{}p{4.0cm}p{3.7cm}@{}}
\toprule
Item & Value \\
\midrule
Receiver volume $\Vtank$ & 1400~L \\
Compressor $\alpha$, $\beta$, $\gamma$ & 0.07~s$^{-1}$, 294.4~MPa$^{-1}$, 0 \\
Compressor rated flow $\Qspec$ & 6800~L/min (plant: 7300~L/min) \\

Fixed-speed compressor flow & 4000~L/min \\
Required pressure $\pneed$ & 0.50~MPa \\
Target pressure $\pref$ & 0.572~MPa \\
Set-pressure bounds & 0.50/0.95~MPa \\
Sampling and control period $\Delta t$ & 1~s \\
Prediction/control horizon $T_p$/$T_c$ & 30~s/1 \\
Weights $w_{\mathrm{err}}$, $w_{\mathrm{term}}$ & 1, $10$ \\
Iteration limit $N_{\mathrm{iter}}$/tolerance & 3/$10^{-3}$ \\
Random base load & uniform 240--360~L/min, seed 2 (200--300~L/min, seed 3 in Section~\ref{subsec:real_demand}) \\

\bottomrule
\end{tabular}
\end{table}

Three types of controller are compared.
\begin{description}
\item[K0 (constant set pressure)] Constant operation at $\pset=0.70$~MPa, the full-load discharge pressure of many plant compressors \citep[p.~32]{DOE_CA_Guide}. A constant set pressure cannot follow the demand and must never let the end-use pressure fall below the limit, so it keeps at all times a margin for the largest drop and for unforeseen events. Section~\ref{subsec:real_demand} also uses the current operation of the target plant.

\item[K1 (end-use pressure PID cascade)] The most standard existing method that a practitioner can add with the same gauge at the end-use point. The outer loop is a discrete PI controller (1-s period, parallel form, output saturation 0.50--0.95~MPa, clamping anti-windup) with input $\pref-\pend$ and output $\pset$; the inner loop is the existing compressor control. Reusing the gains of the built-in inverter PI in the cascade (\textbf{K1-invPI}, $K_c=52.5$, about 28 times the ultimate gain of the end-use pressure loop; Appendix~\ref{app:pid}) makes the set pressure switch between its bounds, so that option is not considered further. \textbf{K1-SIMC} is tuned
 with the Skogestad internal model control (SIMC) rule \citep{Skogestad2003} at its recommended value $\tau_c=\theta$ ($K_c=0.388$, $\tau_I=15.7$~s), based on a first-order-plus-dead-time (FOPDT) model ($K=1.03$, $\tau=15.7$~s, $\theta=19.7$~s) identified from a step test ($\pset$ from 0.60 to 0.65~MPa) under constant load; it represents a PI tuned with a standard rule at its recommended setting (Appendix~\ref{app:pid}).
\item[K2 (proposed NMPC)] The controller of Section~\ref{sec:control}.
\end{description}
For fairness, all controllers use only the same measurements (K1 uses only $\pend$) and share the target $\pref$, the threshold $\pneed$, the output bounds and the control period; parameters are not changed across scenarios; the random seed is the same for all controllers; and the design of K1-SIMC does not use the internal models of the simulator. Prior system-level methods for compressed air systems \citep{Zhong2023,Lassahn2026} optimize the scheduling and loading of the compressors for a given or forecast demand rather than the set pressure against an end-use requirement; they address a different decision problem and are therefore not compared quantitatively.

The metrics over the evaluation interval are the mean set pressure $\bar p_{\mathrm{set}}$, mean receiver pressure $\bar p_{\mathrm{tank}}$, mean end-use pressure $\bar p_{\mathrm{end}}$, minimum end-use pressure $p_{\mathrm{end}}^{\min}$, violation time $T_{\mathrm{viol}}=\sum_t \mathbb{1}[\pend(t)<\pneed]\Delta t$, violation integral $I_{\mathrm{viol}}=\sum_t\max(0,\pneed-\pend(t))\Delta t$, settling time $T_{\mathrm{settle}}$ (the first time after the load is removed at which $|\pend-\pref|<5$~kPa holds and continues to hold until the end of the interval), total variation of the set pressure $TV_{\mathrm{set}}=\sum_t|\pset(t)-\pset(t-\Delta t)|$, number of set-pressure changes $N_{\mathrm{chg}}$ ($|\Delta\pset|>0.005$~MPa), and the estimated annual energy consumption $E_{\mathrm{year}}$ from the power model of Appendix~\ref{app:power}.

\subsection{Evaluation with real demand at the controlled end-use point (main result)}\label{subsec:real_demand}
\paragraph{Input data}
Measurements from two consecutive weekday daytime hours (7,200~s, preceded by 200~s for initialization) at the target plant were used. The receiver pressure and the pressure at the controlled end-use point were measured; since end-use flows are not measured (R2), the flow at the controlled end-use point (E20 in Fig.~\ref{fig:sim_layout}) was reconstructed from the measured pressure drop $\ptank-\pend$. (1) The steady-state pressure drop $\Delta p(Q)$ for an end-use flow of 0--1500~L/min was calibrated from the static behavior of the pipe-network simulator (inlet-pressure dependence corrected by $\Delta p\propto p^{-1.08}$, cubic polynomial fit, root-mean-square error (RMSE) 0.02~kPa). (2) The measured pressure drop was smoothed with a 60-s centered moving average and converted to flow with the inverse of $\Delta p(Q)$. The smoothing is needed because the receiver of the real plant oscillates with an amplitude of about 0.03~MPa and a period of 1--2~min under load/unload operation, and the resulting pipe-filling transients enter the pressure drop; the evaluation thus covers demand changes on time scales of 1~min or longer. The loads at the other 19 end-use points were not measured and were assumed: the 18 base-load points draw uniformly random flows within $\pm 20\,\%$ of 250~L/min (fixed seed), and the neighboring point draws a constant 100~L/min. The reconstructed flow at the controlled point has a mean of 367~L/min and a range of 196--814~L/min (Fig.~S6 of the Supplementary material).

\paragraph{Reproducibility}
An open-loop replay of the pipe-network simulator, driven by the reconstructed flow and the measured receiver pressure, reproduced the measured end-use pressure with an MAE of 3.9~kPa (maximum 17~kPa), similar to the agreement in Section~\ref{subsec:simulator_verification}. Most of the residual is the transient caused by the receiver oscillation, not an error of the demand reconstruction.

\paragraph{Comparison conditions}
Two constant set-pressure cases were used: K0 (0.70~MPa, constant operation with a margin, Section~\ref{subsec:setup}) and K0$'$ (0.62~MPa, slightly below the mean measured receiver pressure of 0.626~MPa, equivalent to the current operation of the target plant). K0$'$ is about the lowest constant set pressure that meets $\pneed$ in the design event of Section~\ref{subsec:artificial} (Section~\ref{subsec:energy}) and so has no margin for larger events.

 Together with K1-SIMC and K2, the four controllers were compared with the target, threshold and bounds of Section~\ref{subsec:setup}. The evaluation interval is 0--7200~s.

\begin{figure*}[pos=!tp]
 \centering
 \includegraphics[width=\linewidth]{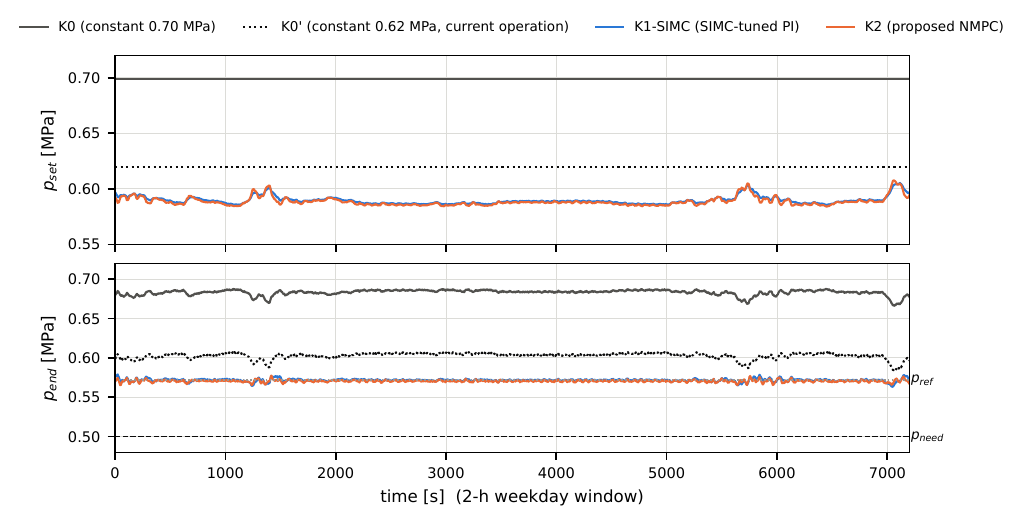}
 \caption{Closed-loop comparison with the real demand time series at the controlled end-use point. Top: set pressure; bottom: end-use pressure. K0 is constant at 0.70~MPa and K0$'$ at 0.62~MPa (equivalent to the current operation). Dashed line: $\pneed$; dotted line: $\pref$.}
 \label{fig:real_demand}
\end{figure*}

\paragraph{Results}
Fig.~\ref{fig:real_demand} shows the time series and Table~\ref{tab:real_demand} the metrics. No controller violated the required pressure. The proposed K2 kept the set pressure at 0.589~MPa and the receiver pressure at 0.588~MPa on average, and held the end-use pressure near the target of 0.572~MPa (mean 0.571~MPa, standard deviation 1.4~kPa, minimum 0.565~MPa). The estimated annual energy consumption is 9.3\,\% lower than with K0 and 2.8\,\% lower than with K0$'$. K1-SIMC also reached a set-pressure level equivalent to K2 without violation (mean 0.590~MPa, 9.2\,\% saving), with a standard deviation of the end-use pressure of 1.7~kPa and a minimum of 0.563~MPa. Under this real demand, the difference between K2 and K1-SIMC is therefore small. The reconstructed demand changes slowly (60-s smoothing, pressure-drop variation of about $\pm 10$~kPa), so the PI cascade can follow it; the advantage of predictive, anticipatory action appears when demand changes abruptly (Section~\ref{subsec:artificial}). The NMPC converged within the iteration limit in 90\,\% of all cycles and applied the intermediate solution in the rest; there was no cycle without a feasible solution.

K0$'$ is an idealized operation that holds the mean pressure of the actual operation; the simulated compressors have no load/unload losses for any controller, so the saving relative to the current operation reflects the effect of the set pressure alone.

\begin{table*}[!tp]
\caption{Closed-loop metrics under real demand (evaluation interval 0--7200~s, pressures in MPa gauge).}
\label{tab:real_demand}
\centering
\footnotesize
\begin{tabular}{lcccccccc}
\toprule
Controller & $\bar p_{\mathrm{set}}$ & $\bar p_{\mathrm{tank}}$ & $\bar p_{\mathrm{end}}$ & $p_{\mathrm{end}}^{\min}$ & $T_{\mathrm{viol}}$ [s] & $TV_{\mathrm{set}}$ [MPa] & $N_{\mathrm{chg}}$ & $E_{\mathrm{year}}$ [kWh] (vs.\ K0/K0$'$) \\
\midrule
K0 (constant 0.70) & 0.699 & 0.698 & 0.683 & 0.666 & 0 & 0 & 0 & 345,069 (---/$+7.1$\,\%) \\
K0$'$ (constant 0.62, current operation) & 0.620 & 0.620 & 0.603 & 0.584 & 0 & 0 & 0 & 322,233 ($-6.6$\,\%/---) \\
K1-SIMC (SIMC tuning) & 0.590 & 0.589 & 0.572 & 0.563 & 0 & 0.65 & 0 & 313,416 ($-9.2$\,\%/$-2.7$\,\%) \\
K2 (proposed) & 0.589 & 0.588 & 0.571 & 0.565 & 0 & 0.76 & 0 & 313,135 ($-9.3$\,\%/$-2.8$\,\%) \\
\bottomrule
\end{tabular}
\end{table*}

\subsection{Evaluation with synthetic load}\label{subsec:artificial}
To compare the responses to sudden increases and decreases in demand clearly, a synthetic load was also used. The flow at the controlled end-use point is trapezoidal: it rises from a steady 100~L/min to 1200~L/min, the step for which the margin $\pref-\pneed$ is designed, at $t=450$~s
 and returns to 100~L/min at $t=600$~s. To evaluate pressure interaction through nearby pipes, the neighboring end-use point receives a load of the same shape (100 to 1200~L/min) during $t=200$--800~s. All changes are 10-s linear ramps. The other 18 points draw uniformly random flows of 240--360~L/min. The evaluation interval is 0--1000~s.

\begin{figure}[pos=!tbp]
 \centering
  \includegraphics[width=\linewidth]{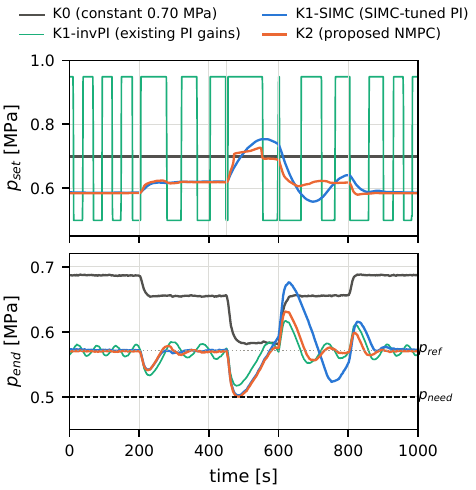}
 \caption{Closed-loop comparison under synthetic load. Top: set pressure; bottom: end-use pressure. Vertical lines mark the start and end of the load at the neighboring end-use point (200--800~s) and at the controlled end-use point (450--600~s). K1-invPI (built-in PI gains reused, Section~\ref{subsec:setup}) is shown for reference. Dashed line: $\pneed$; dotted line: $\pref$.}
 \label{fig:compare_art}
\end{figure}

Fig.~\ref{fig:compare_art} shows the set pressure and end-use pressure of each controller, and Table~\ref{tab:artificial} the metrics. With K0 the end-use pressure always exceeds the required pressure by a wide margin (minimum 0.581~MPa). The proposed K2 lowers the set pressure to near its lower bound while the loads at the neighboring and controlled points are low. When the consumption rises sharply around $t=200$~s and 450~s, the demand estimate of Eq.~\eqref{eq:demand_estimate} rises with the consumption, the receiver model predicts the resulting fall of the receiver pressure, and K2 raises the set pressure in advance; the held pressure drop is updated every cycle as it grows.
 When the load returns, K2 lowers the set pressure again on its own. The mean set pressure is 0.617~MPa with K2 against 0.699~MPa with K0, about 0.08~MPa lower, and the minimum end-use pressure is 0.5002~MPa, meeting the required 0.50~MPa (no violation).

K1-SIMC operates smoothly without violation but tracks the transients worse than K2: after the load is removed, the integral lag lets the end-use pressure overshoot to 0.68~MPa and then undershoot to 0.52~MPa before it settles. With the same sensors, a SIMC-tuned PI can thus also lower the set pressure, while K2, with its anticipatory action, performs better in the transients. K1-invPI (Section~\ref{subsec:setup}) switches the set pressure between its bounds throughout, as its gain is far above the ultimate gain.

\begin{table*}[!tp]
\caption{Closed-loop metrics under synthetic load (evaluation interval 0--1000~s, pressures in MPa gauge). Settling is the time after the load is removed at which $|\pend-\pref|<5$~kPa is reached (---: not reached).}
\label{tab:artificial}
\centering
\footnotesize
\begin{tabular}{lccccccccc}
\toprule
Controller & $\bar p_{\mathrm{set}}$ & $\bar p_{\mathrm{tank}}$ & $\bar p_{\mathrm{end}}$ & $p_{\mathrm{end}}^{\min}$ & $T_{\mathrm{viol}}$ [s] & $TV_{\mathrm{set}}$ [MPa] & $N_{\mathrm{chg}}$ & Settling [s] & $E_{\mathrm{year}}$ [kWh] (vs.\ K0) \\
\midrule
K0 (constant 0.70) & 0.699 & 0.698 & 0.657 & 0.581 & 0 & 0 & 0 & --- & 400,904 (---) \\
K1-SIMC (SIMC tuning) & 0.619 & 0.617 & 0.572 & 0.502 & 0 & 0.56 & 0 & 910 & 379,856 ($-5.3$\,\%) \\
K2 (proposed) & 0.617 & 0.612 & 0.567 & 0.500 & 0 & 0.41 & 5 & 875 & 376,087 ($-6.2$\,\%) \\
\bottomrule
\end{tabular}
\end{table*}

Over all control cycles of the evaluation interval, the NMPC found a solution satisfying the optimality condition within the iteration limit in 71\,\% of the cycles; in 29\,\% it reached the limit $N_{\mathrm{iter}}=3$ and applied the intermediate solution, and in one cycle (0.1\,\%) the model equations were not met within the tolerance and the previous set pressure was kept.
 Since the control performance above was obtained under these conditions, the small iteration limit is sufficient in practice.

\subsection{Robustness to facility changes (plant expansion and network split)}\label{subsec:facility_change}
\begin{figure}[pos=!tbp]
 \centering
 \includegraphics[width=\linewidth]{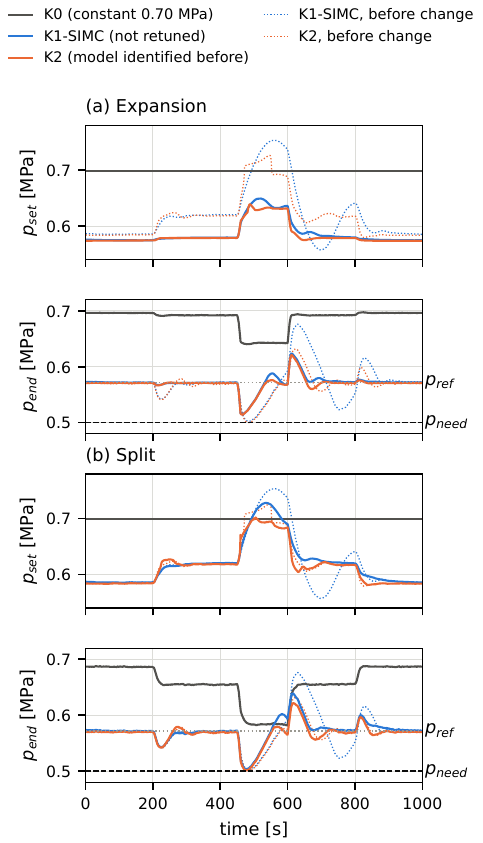}
 \caption{Robustness to facility changes: (a) plant expansion (B1), (b) network split (B2). For each case, the upper panel shows the set pressure and the lower panel the end-use pressure. Solid lines: K1-SIMC with its initial tuning (``not retuned'') and K2 with its initial model (``model identified before'') after the change; thin dotted lines (``before change''): the nominal network (Section~\ref{subsec:artificial}). Dashed line: $\pneed$; dotted line: $\pref$.}
 \label{fig:facility_change}
\end{figure}

This section tests R4: whether controllers identified and tuned on the nominal network can be used unchanged after a facility change. Two changes were applied to the pipe-network simulator, and each run started from steady state on the changed network (changing pipes during operation is not simulated).
\begin{enumerate}[(B1)]
\item Plant expansion: the supply pipes to the controlled point (P10 and P11, 134~m in total) are widened 1.5 times (27.6 to 41.4~mm) and the main sections near it (P12\_2 and P33\_1) 2 times (67.9 to 135.8~mm), the receiver is doubled to 2800~L, and the fixed-speed compressor is enlarged from 4000 to 6800~L/min. The flow at the controlled point rises 1.6 times (steady 160~L/min, 1900~L/min during the step).
\item Network split: the target network is split off with half the receiver (700~L), and its branch to the controlled point (P33\_4) is narrowed from 16.1 to 12.9~mm, as by a half-closed valve. The split is assumed to accompany a change in production that lowers the step at the controlled point to 930~L/min, which gives about the same pressure drop as the design event, so the margin designed for that event still applies.
\end{enumerate}
The other loads are as in Section~\ref{subsec:artificial}. K0 (constant 0.70~MPa) is compared with K1-SIMC with its initial tuning and K2 with its initial end-use model and weights (``not retuned'' and ``model identified before'' in Fig.~\ref{fig:facility_change}). K2 is given the new receiver volume as a nameplate value (the fixed-speed flow is not a parameter of K2), whereas K2-noupd keeps the initial 1400~L.

Both controllers coped with both changes without retuning (Fig.~\ref{fig:facility_change}, Table~\ref{tab:facility_change}): no violation, and minimum end-use pressures at least as high as on the nominal network (0.501--0.515 against 0.500--0.502~MPa). K2 settled earlier than K1-SIMC in both cases (691 against 815~s, 868 against 898~s) and saved as much or more energy (7.4 and 7.4\,\%, 6.7 and 6.2\,\% relative to K0); after the split its set pressure varied slightly more (0.40 against 0.34~MPa). The expansion lowered both set pressures (0.587 and 0.584~MPa), because the larger supply pipes almost remove the pressure drop from the nearby loads.

K2 needs no update because both changes shift the pressure-drop level, which the held measured drop absorbs (Section~\ref{subsec:multistep}): with its initial model, its one-step error was 0.55 and 0.90~kPa after the changes, against 0.91~kPa on the nominal network. Re-identification on K0 data (1000~s) after each change returned almost the initial model (sign bounds, $a_{\mathrm{end}1}=a_{\mathrm{end}2}\approx 0$, $b_{\mathrm{end}1}=-b_{\mathrm{end}2}=1$; one-step MAEs 0.27 and 0.38~kPa), a bound solution that the check of Section~\ref{subsec:reidentification} flags as too little excitation. Even the receiver volume need not be updated: with the initial 1400~L, half or twice the true value (K2-noupd), the metrics hardly changed, because the volume enters both Eqs.~\eqref{eq:tank_dynamics} and \eqref{eq:demand_estimate} and its error cancels in the current rate of pressure change.

These changes did not challenge the PI either, whose SIMC gain has a margin of about 5 to the ultimate gain (Appendix~\ref{app:pid}); the two differ in what their tuning requires (Section~\ref{subsec:pid_tuning}). Changes of the dynamics themselves, which re-identification is for, are not produced by this simulator (Section~\ref{subsec:limitations}).

\begin{table*}[!tp]
\caption{Closed-loop metrics for the facility changes (0--1000~s, MPa gauge); nominal network: Table~\ref{tab:artificial}. K2-noupd keeps the initial receiver volume of 1400~L. Settling as in Table~\ref{tab:artificial}. Energy is comparable only within each case, because the expansion changes the fixed-speed compressor.}
\label{tab:facility_change}
\centering
\footnotesize
\setlength{\tabcolsep}{4pt}
\begin{tabular}{llccccccc}
\toprule
Case & Controller & $\bar p_{\mathrm{set}}$ & $p_{\mathrm{end}}^{\min}$ & $T_{\mathrm{viol}}$ [s] & $TV_{\mathrm{set}}$ [MPa] & $N_{\mathrm{chg}}$ & Settling [s] & $E_{\mathrm{year}}$ [kWh] (vs.\ K0) \\
\midrule
(B1) Expansion & K0 & 0.699 & 0.640 & 0 & 0 & 0 & --- & 316,924 (---) \\
 & K1-SIMC & 0.587 & 0.515 & 0 & 0.20 & 0 & 815 & 293,486 ($-7.4$\,\%) \\
 & K2 (proposed) & 0.584 & 0.514 & 0 & 0.18 & 2 & 691 & 293,546 ($-7.4$\,\%) \\
 & K2-noupd & 0.584 & 0.515 & 0 & 0.21 & 6 & 692 & 293,268 ($-7.5$\,\%) \\
\midrule
(B2) Split & K0 & 0.699 & 0.582 & 0 & 0 & 0 & --- & 399,348 (---) \\
 & K1-SIMC & 0.620 & 0.503 & 0 & 0.34 & 0 & 898 & 374,533 ($-6.2$\,\%) \\
 & K2 (proposed) & 0.615 & 0.501 & 0 & 0.40 & 7 & 868 & 372,769 ($-6.7$\,\%) \\
 & K2-noupd & 0.616 & 0.501 & 0 & 0.43 & 10 & 866 & 373,637 ($-6.4$\,\%) \\
\bottomrule
\end{tabular}
\end{table*}

\subsection{PID tuning sensitivity and the difficulty of tuning in the field}\label{subsec:pid_tuning}
To see how much K1-SIMC depends on the recommended value $\tau_c=\theta$, K1-SIMCh ($\tau_c=\theta/2$) and K1-SIMC0 ($\tau_c=0$), which reduce only the closed-loop time constant and use the same step test, were evaluated before the change, after expansion and after the split (Table~S1 of the Supplementary material).

Reducing $\tau_c$ degraded the performance in every condition: the energy reduction fell by 0.2--0.8 points with $\tau_c=\theta/2$ and by 2.3--3.6 points with $\tau_c=0$, and the total variation of the set pressure grew 1.3--3.8 times. Because the loop is dead-time dominant ($\theta=19.7$~s against $\tau=15.7$~s), a higher gain adds overshoot rather than speed, and a PID matches K2 only near the recommended value.

This sensitivity makes tuning in a real plant difficult. The SIMC gain is inversely proportional to the estimated dead time: underestimating $\theta$ by 25\,\% or 50\,\% gives the gains of K1-SIMCh and K1-SIMC0, and even for a noise-free simulated step the estimate differed by 27\,\% between first- and second-order models (Appendix~\ref{app:pid}). In a running plant the load does not stay constant for the $\theta+4\tau\approx 80$~s that a test needs (the pressure drop varied by about $\pm 10$~kPa under real demand), downward steps risk violating $\pneed$, and the tests must be repeated after every facility change.

K2, in contrast, used the same default weights in all scenarios and models identified with Algorithm~1 (Section~\ref{subsec:dataset}), performed comparably to K1-SIMC, and kept this performance after expansion and split, even with the receiver volume off by a factor of two (Section~\ref{subsec:facility_change}). Its merit lies less in the energy saved beyond K1-SIMC than in reaching that performance without a constant-load step test at the site or a choice of $\tau_c$, and keeping it after facility changes; its compressor model needs one set-pressure change per compressor product under any load (Section~\ref{subsec:dataset}), not a test after each change.

\subsection{Execution time on the industrial controller}\label{subsec:runtime}
To confirm that the NMPC can be implemented in practice, the control algorithm (demand estimation, forward simulation of the three models, analytic Jacobians and SQP iterations) was implemented in C on the industrial controller used in actual sequencing controllers. The computation time from receiving the measurements to computing the set pressure was measured, excluding communication and logging. The inputs covered 100 cases over the operating range (pressures, flows, receiver volume and set-pressure bounds; Section~S10 of the Supplementary material);
 each case was run three times (300 trials). The controller has 16~MB of program memory, 16~MB of (volatile) data memory and a minimum execution time of 6.6~ns for double-precision floating-point instructions.

The computation time was 0.7109~s at minimum, 1.4125~s at maximum and 1.0827~s on average, in two levels (Fig.~S7 of the Supplementary material), 0.71--0.72~s (138 trials) and 1.37--1.41~s (162 trials); repeating a case changed it by at most 0.031~s, so it is determined by the input conditions. It was measured with the computation alone; the slower level, about twice the faster, probably reflects more SQP iterations. In operation the NMPC is to run beside the existing control tasks at a lower priority.
 Because $N_{\mathrm{iter}}=3$ bounds the computation per cycle by $O(T_p N_{\mathrm{iter}})$ (Section~\ref{subsec:computation}), the time is not only short but also bounded in advance.

Because the computation time is bounded, the update period can be chosen independently of the 1-s model sampling; applying each result as soon as it is computed gives a period of at most 1.4125~s, roughly 2~s with communication, which is one third of the time the discharge flow needs to follow a set-pressure change (about 6~s) and short compared with the 35-s rise of the pressure drop in a sudden demand increase (Section~\ref{subsec:deployment}). With updates every 2~s applied 1.4~s late, K2 under the synthetic load reached 0.498~MPa (17~s and about 2~kPa below $\pneed$) and used 0.07\,\% more energy.
 R3 is therefore met with the 1-s update used here, while a 2-s update needs a margin about 2~kPa larger.

\subsection{Estimated energy saving}\label{subsec:energy}
To convert the effect into energy, the annual consumption of an assumed station with a fixed-speed compressor for the base load and a variable-speed one for tracking was estimated with the power model of Appendix~\ref{app:power}, which assumes power linear in the load factor, 8.4\,\% less power per 0.1~MPa lower discharge pressure and 6000 operating hours per year; it is an estimate, not a measurement.

Under the synthetic load of Section~\ref{subsec:artificial}, K2 consumes 376,087~kWh per year against 400,904~kWh with K0, a reduction of 24,817~kWh (6.2\,\%). K1-SIMC reduces it by 5.3\,\% (Table~\ref{tab:artificial}). It is smaller than the 6.9\,\% that the pressure coefficient alone gives for the mean reduction of about 0.08~MPa, because set-pressure peaks coincide with high discharge flow (Appendix~\ref{app:power}). In this interval, dominated by the design event, the lowest constant set pressure that just survives it (about 0.62~MPa; the K0 run lowered by 0.08~MPa reaches about 0.50~MPa) would save about 6.6\,\%, slightly more than K2. That value, however, is known only after the event and leaves no allowance for any larger one, so it is not an operable setting where the lower limit must never be violated; K2 keeps its margin on the end-use pressure instead, raises the set pressure only when the demand rises, and can raise it up to 0.95~MPa for larger events.
 Under the real demand time series (Section~\ref{subsec:real_demand}), K2 consumes 313,135~kWh against 345,069~kWh with K0, a reduction of 31,934~kWh (9.3\,\%), larger than under the synthetic load because the real demand was lower and the margin of constant-pressure operation more excessive. This is where the dynamic set pressure pays off: K2 uses 9,098~kWh (2.8\,\%) less than K0$'$ (322,233~kWh), which holds the 0.62~MPa needed for the design event at all times without any further margin. Evaluating the pressure factor with the receiver pressure instead of the set pressure changes the savings of K2 and K1-SIMC by at most 0.5 points; with a pressure coefficient of 7.3\,\%/0.1~MPa \citep{DOE_CA_Guide}, the savings of K2 are 5.3\,\% (synthetic load) and 8.0\,\% (real demand), and 2.4\,\% relative to K0$'$.

\section{Discussion}\label{sec:discussion}

\subsection{What the staged validation shows and does not show}\label{subsec:staged_discussion}
The staged validation (Section~\ref{subsec:staged}) checks model validity, closed-loop behavior and feasibility on the target hardware, each with the most realistic means available without operational risk. Stage (i) showed that the constrained identification yields models usable in the NMPC from real plant data (one-step errors of 1.1\,\% for the discharge flow and 0.2 and 1.5~kPa for the receiver and end-use pressures), whereas unconstrained identification with the same regressors does not when the pressure drop is held as in the NMPC.
 Stage (ii) showed, on a simulator that agrees with measurements within 5--8~kPa, that the method lowers the set pressure while meeting the required pressure, is comparable to a SIMC-tuned PI cascade under real demand at the controlled end-use point and keeps its performance after facility changes, whereas the PID performance depends strongly on $\tau_c$. Stage (iii) showed that the actual controller hardware updates the set pressure well within the compressor response time. What only closed-loop trials in the plant (stage (iv), Section~\ref{subsec:deployment}) can confirm is stated in the scope of the evaluation below.

\subsection{Scope and limitations}\label{subsec:limitations}
\textbf{Scope of the method.} The method serves a centralized compressor station through one representative critical end-use point (P5); several points can be handled as in Section~\ref{subsec:deployment}. The required pressure is secured not by a state constraint but by the margin $\pref-\pneed$, a design parameter sized from plant data and the update period (Sections~\ref{subsec:runtime} and~\ref{subsec:deployment}; here 0.072~MPa for the 1200-L/min design event); a drop that outruns the pressurization of the receiver is beyond any set-pressure control. Because the demand estimate treats the receiver as the only storage, the set pressure stays high briefly after a sudden load increase (about 10~s in the design event, about 150~s at 1500~L/min), on the safe side; an effective storage volume identified from operating data is expected to remove this. Robustness is shown by the facility changes and a twofold error in the receiver volume (Section~\ref{subsec:facility_change}), not by a stability-margin analysis.

\textbf{Scope of the evaluation.} The simulator agrees with measurements within 5--8~kPa (3.9~kPa in the replay of real demand), so only larger differences between controllers are interpreted, and K2 and K1-SIMC are comparable under real demand. The baseline is the PID cascade with the same added gauge; baselines that isolate the effect of prediction are not included. The simulator idealizes the plant: K2 uses its compressor and receiver models, the measurements are exact and undelayed apart from the 2-s update of Section~\ref{subsec:runtime}, the station has no minimum speed, unloading or start/stop (surplus supply after load removal is taken by a negative variable-speed flow, without effect on the minimum end-use pressures), and the network shows practically no damped oscillation and changes only its pressure-drop level under pipe and receiver changes. These effects, the load/unload cycling of the plant (about 0.03~MPa, period 1--2~min), re-identification after a real facility change, operation over months and the actual energy consumption are left to stage (iv). Within the simulation, the real demand is reconstructed at the controlled point (60-s smoothing, one two-hour interval) with assumed loads elsewhere, and the energy saving is estimated with the power model of Appendix~\ref{app:power}, without load/unload losses and with pressure-independent demand, so leaks would add to it. With the faster compressor of the plant, K2 and a SIMC PI retuned for it performed alike (Table~S2 of the Supplementary material), so the transient advantage of K2 in Section~\ref{subsec:artificial} depends on the slower compressor of the simulator. The compressor model rests on one upward set-pressure change, whereas the end-use model kept one-step errors of 1.04--1.17~kPa when identified on each of the 13 data files, with poles of 0--0.75, provided $a_{\mathrm{end}1}<0$ (Table~S3).

\subsection{Deployment and implications for supply-system operation}\label{subsec:deployment}
The method is transferred as a retrofit: another plant needs one added pressure gauge at the critical end-use point and access to the compressor data (Table~\ref{tab:signals}), identification with Algorithm~1 from a few days of operating data, the controller of Section~\ref{sec:control} on the sequencing controller and the procedure of Section~\ref{subsec:reidentification}; none requires a piping survey or flow meters.

Stage (iv) introduces the controller under supervision and removes operational risk step by step: the set pressure is applied with its lower bound near the current operating pressure, and the bound is widened gradually while the end-use pressure margin is checked.
 Because only the set pressure is changed, resetting it to the conventional constant value on a sensor fault, a prediction error above a threshold or a solver failure returns the plant directly to the existing operation; the lower bound of the set pressure would be set to $\pneed$ plus the largest pressure drop in operation (0.58~MPa with the 80-kPa maximum below, under the current 0.62~MPa) rather than to the mechanical limit, with a rate limit on lowering.

Plants with several critical end-use points can be handled by identifying the model of Section~\ref{subsec:end_model} for each point and using the minimum over the points of the difference between predicted and required pressure in the objective, keeping a single decision variable.

Beyond lowering the set pressure, the controller quantifies how much headroom the supply system has. The margin $\pref-\pneed$ is set by the time scales of a sudden demand increase rather than by the control law. When the flow at the controlled point rose to 1200~L/min in 10~s (Section~\ref{subsec:artificial}), the pressure drop grew by about 0.08~MPa in about 35~s, so the end-use pressure fell at about 2~kPa/s, whereas the receiver can be pressurized only as fast as the spare compressor capacity divided by the storage volume allows: 0.34~kPa/s on average during those 35~s and at most about 0.85~kPa/s, because the pipes (about 1.4~m$^3$) and an air reservoir node (2.26~m$^3$, Table~S6) add to the 1.4-m$^3$ receiver and the variable-speed compressor follows a set-pressure change with a time constant of about 14~s.
 The margin covered the remaining drop, and the minimum end-use pressure was 0.500~MPa with K2 and 0.502~MPa with K1-SIMC. Constant-pressure operation passes such a change only because it always holds a large margin, which is exactly its energy loss in normal operation. The margin is therefore a design parameter that trades energy saving against headroom; it can be sized from (a) the largest expected pressure-drop increase, such as a simultaneous start of critical end-use equipment, and (b) the ratio of the rate of pressure fall at the end-use point to the rate of pressurization. In the plant data, the drop to P5 rose by at most 55~kPa within 30~s and reached at most 80~kPa, about the drop of the design event.
 The real demand of the target plant (about $\pm 10$~kPa on time scales of 1~min or longer) stayed within this headroom.

Loads beyond this headroom point to the piping and the compressor capacity rather than to the controller: with a step of 1500~L/min, which saturated the compressors, and of 2800~L/min after adding compressors, both K1-SIMC and K2 fell below $\pneed$ (77--84~s, minimum 0.45--0.47~MPa), and only constant 0.70~MPa operation avoided it.
 These loads, like the 1200-L/min design event chosen as a severe test, exceed what a properly designed system carries, a pressure loss from the compressor discharge to the point of use much less than 10\,\% of the discharge pressure \citep[p.~31]{DOE_CA_Guide}, here 70~kPa: the steady drop to the controlled point is 128~kPa at 1200~L/min and 126--163~kPa for these steps, against 35~kPa at the largest real demand.
 Because the controller observes $\pdiff$ and $\hat Q_{\mathrm{tank}}$ at every cycle, it detects such loads in operation and shows where an enlarged branch pipe, a local receiver near the critical end-use point or feedforward of known demand changes would let the set pressure be lowered further; its signals thus inform the operation of the whole supply system, not only its set pressure.

\section{Conclusions}\label{sec:conclusion}
End-use-aware set-pressure control can be retrofitted to a centralized compressor station within the four practical constraints of Section~\ref{subsec:background}: without a piping survey or flow meters, on the hardware of existing sequencing controllers and without retuning after facility changes. Validated in stages on plant data, a pipe-network simulator checked against measurements and the actual controller hardware, the NMPC lowered the set pressure by about 0.03~MPa below the current operation, which already runs at about the lowest constant set pressure that covers the design event, while meeting the required end-use pressure; the estimated annual energy consumption falls by 2.8\,\% (6--9\,\% relative to the common 0.70~MPa). It needed one added pressure gauge, kept the required pressure with its initial model and default weights after plant expansion and network split, and computed each update in at most 1.41~s, well below the compressor response time of 6~s.

The methodological core is the role of the identification constraints. Constraints that build the known pressure response of a pipe network into least squares (damped oscillation, stability and a constant-drop equilibrium) do not make the model more accurate; they make it usable in the way a predictive controller uses it. With the measured pressure drop held over the horizon, the constrained model is stable and passes the held drop through with unit gain, whereas the same regressors identified without constraints diverge and a standard ARX model amplifies its held error eightfold. The constraints thus yield offset-free prediction in which the held disturbance is a measured physical quantity rather than an estimated one, and they make robustness to facility changes a property of the model structure: a change in the pressure-drop level is absorbed by the measurement, not by retuning. Choosing identification constraints for the way a model is used in the controller rather than for its fit may serve other networks in which a measured pressure difference between a supply point and a consumer carries an unmeasured load, such as industrial gas headers and water distribution networks.

For industry, the method turns set-pressure reduction from a site-specific engineering project into a retrofit. The investment is one pressure gauge and software on sequencing controllers already installed; the models follow by a fixed procedure from a few days of operating data and one set-pressure change per compressor product; and no constant-load step test at the site is needed, whereas a PID cascade with the same gauge reaches a comparable saving only when tuned from such a test, which is difficult in a running plant. The centralized station it targets is the default layout in industrial practice. It also changes how supply margins are managed: the margin that constant-pressure operation holds at all times becomes an explicit design parameter $\pref-\pneed$ sized from plant data, and the signals of the controller show where loads exceed the design capacity of the branch piping. The few signals that the controller already reads thus inform the operation of the whole supply system: they decide the set pressure at each cycle and show the margin to keep, the branches to enlarge and the local receivers to add.

\appendix
\setcounter{equation}{0}
\renewcommand{\theequation}{\thesection.\arabic{equation}}
\section{Power model}\label{app:power}
The fixed-speed and variable-speed compressors both have a rated power of 42.55~kW and a rated maximum flow of $\Qspec=6800$~L/min. The fixed-speed compressor runs at a constant 4000~L/min, and the variable-speed compressor supplies the remaining demand. With an overall efficiency of 0.943, the power at rated load is $P_{\mathrm{load}}=42.55/0.943\approx 45.1$~kW, and the unload power is $P_{\mathrm{unload}}=0.10P_{\mathrm{load}}$. The base power of the variable-speed compressor is interpolated linearly in the discharge flow:
\begin{align}
 P_{\mathrm{vsd,raw}}(t) &= \frac{\Qcomp(t)}{\Qspec}\left(P_{\mathrm{load}}-P_{\mathrm{unload}}\right)+P_{\mathrm{unload}}\\
 &\qquad P_{\mathrm{unload}}\le P_{\mathrm{vsd,raw}}(t)\le P_{\mathrm{load}} \nonumber\\
 F_{\mathrm{press}}(t) &= 1-\frac{p_{\mathrm{spec}}-\pset(t)}{0.1}\times 0.084\\
 P_{\mathrm{vsd}}(t) &= P_{\mathrm{vsd,raw}}(t)\,F_{\mathrm{press}}(t)\\
 P_{\mathrm{fixed}}(t) &= P_{\mathrm{load}}\,F_{\mathrm{press}}(t)\\
 P_{\mathrm{total}}(t) &= P_{\mathrm{vsd}}(t)+P_{\mathrm{fixed}}(t)
\end{align}
with $p_{\mathrm{spec}}=0.7$~MPa. The annual energy consumption is the mean of $P_{\mathrm{total}}$ over the evaluation interval multiplied by 6000 operating hours per year. Neglecting the saturation of $P_{\mathrm{vsd,raw}}$, $\mathrm{E}[P_{\mathrm{vsd}}]=(a\,\mathrm{E}[\Qcomp]+P_{\mathrm{unload}})\,\mathrm{E}[F_{\mathrm{press}}]+0.84\,a\,\mathrm{Cov}(\Qcomp,\pset)$ with $a=(P_{\mathrm{load}}-P_{\mathrm{unload}})/\Qspec$, so set-pressure peaks that coincide with high discharge flow cost more than their share of the mean set pressure, whereas the fixed-speed compressor depends only on the mean set pressure.

\setcounter{equation}{0}
\section{Details of the PID cascade}\label{app:pid}
The outer loop uses the discrete parallel form
\begin{align}
 u(k)&=K_c\Big[e(k)+\frac{\Delta t}{\tau_I}\sum_{j\le k}e(j) \nonumber\\
 &\qquad +\tau_D\frac{e_f(k)-e_f(k-1)}{\Delta t}\Big]
\end{align}
with $e=\pref-\pend$, output saturation 0.50--0.95~MPa (the same bounds as K2), clamping anti-windup, an initial output of 0.70~MPa (as K0, with bumpless initialization of the integrator) and a derivative filter discretized by backward Euler. All parameter sets use $\tau_D=0$ (PI).

Conversion of the built-in PI gains (Section~\ref{subsec:setup}): a proportional gain of 50\,\%/\% on a 0--1-MPa sensor, with 100\,\% output mapped to the set-pressure range of 1.05~MPa, gives $K_c=50\times 1.05/1.0=52.5$~MPa/MPa (integral time 0.10~s).
Identification for K1-SIMC: under constant load (base load 300~L/min, controlled and neighboring points 100~L/min, no randomness), $\pset$ was stepped from 0.60 to 0.65~MPa and the response was recorded for 300~s. Fitting an FOPDT model gave $K=1.03$, $\tau=15.7$~s and $\theta=19.7$~s (RMSE 0.0025~MPa). Fitting a second-order-plus-dead-time model to the same response gave $\tau_1=\tau_2=10.3$~s and $\theta=14.3$~s (RMSE 0.0021~MPa, a 16\,\% improvement), so the dead-time estimate differed by 27\,\% depending on the model form. The FOPDT model, which the SIMC rule assumes, was adopted. The SIMC rule \citep{Skogestad2003} $K_c=\tau/(K(\tau_c+\theta))$, $\tau_I=\min(\tau,4(\tau_c+\theta))$ with $\tau_c=\theta$ gives $K_c=0.388$ and $\tau_I=15.7$~s. For the sensitivity study, K1-SIMCh ($\tau_c=\theta/2$) has $K_c=0.517$ and K1-SIMC0 ($\tau_c=0$) has $K_c=0.776$, both with $\tau_I=15.7$~s. The ultimate gain of the end-use pressure loop under proportional control is about 1.9.

\section*{Acknowledgements}
We thank Tatsurou Yashiki and Katleya Medrano (Hitachi, Ltd.) for providing the measurements and simulation results in Fig.~\ref{fig:simulator_verification}, and Toshiaki Yabe (Hitachi Industrial Equipment Systems Co., Ltd.) for support of the plant measurements.

\section*{Declaration of competing interest}
All authors are employees of Hitachi, Ltd. or Hitachi Industrial Equipment Systems Co., Ltd., which makes air compressors and sequencing controllers. The target plant and its compressors and sequencing controller belong to Hitachi Industrial Equipment Systems, the simulator is described in a patent of Hitachi Industrial Equipment Systems, the coefficient of 8.4\,\% is taken from a document of Hitachi Industrial Equipment Systems, and the authors' companies have filed a patent application related to the proposed method. The authors declare no other competing interests.

\section*{Funding}
This research did not receive any specific grant from funding agencies in the public, commercial, or not-for-profit sectors.

\section*{Data availability}
The plant data and the simulator are proprietary to the authors' companies and cannot be shared; the parameters, settings and algorithm are in the paper, and the data sources, scenario settings and pipe-network parameters in Tables~S4--S6 of the Supplementary material.

\section*{Declaration of generative AI and AI-assisted technologies in the manuscript preparation process}
During the preparation of this work the authors used Claude Opus 5.5 (Anthropic) to organize the manuscript structure, draft and revise the text, assist translation into English, write the code described in Section~\ref{subsec:staged} and draw the explanatory figures. After using this tool, the authors reviewed and edited the content as needed and take full responsibility for the content of the published article.

\printcredits

\bibliographystyle{cas-model2-names}
\bibliography{refs}

\clearpage
\makeatletter
\edef\arxivmainlastpage{\the\numexpr\value{page}-1\relax}
\AtEndDocument{\immediate\write\@auxout{\string\csxdef{lastpage}{\arxivmainlastpage}}}
\makeatother
\includepdf[pages=-]{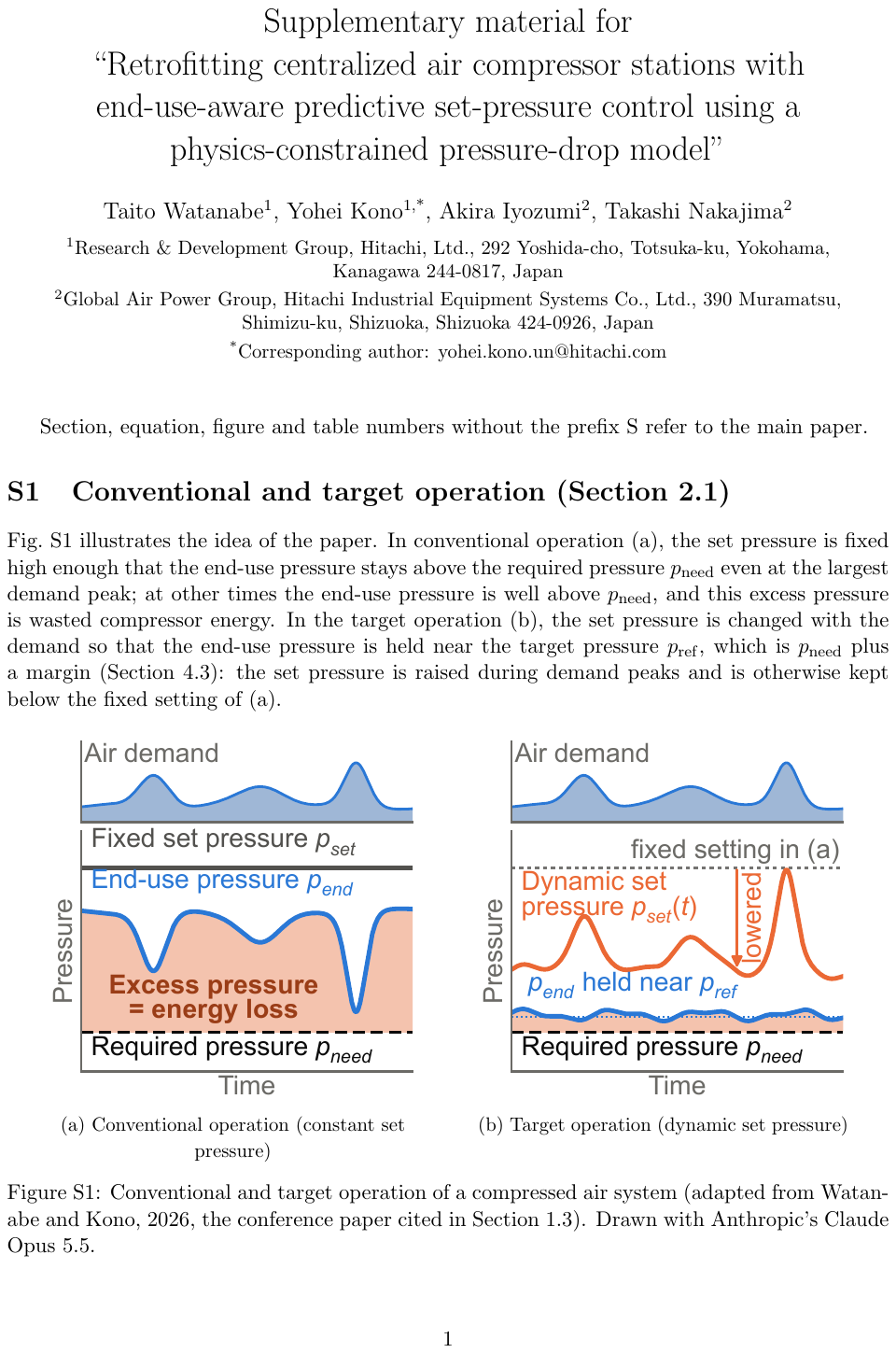}

\end{document}